\documentclass[useAMS,usenatbib]{mn2e}

\usepackage{amsmath}
\usepackage{graphicx}
\usepackage{color}
\usepackage{aas_macros}

\title{Importance of the Initial Conditions for Star Formation~I: Cloud Evolution and Morphology}
\author[Girichidis et al.]{Philipp~Girichidis$^{1,2}$, Christoph~Federrath$^{1,3}$, Robi~Banerjee$^1$, \& Ralf~S.~Klessen$^1$
\vspace*{0.2cm} \\
\scriptsize
$^1$Zentrum f\"ur Astronomie der Universit\"at Heidelberg, Institut f\"ur Theoretische Astrophysik, Albert-Ueberle-Str.~2, 69120 Heidelberg, Germany \\
\scriptsize
$^2$Cardiff School of Physics and Astronomy, The Parade, Cardiff, CF24 3AA, UK \\
\scriptsize
$^3$Ecole Normale Sup\'{e}rieure de Lyon, CRAL, 69364 Lyon Cedex 07, France
}

\newcommand{\ColorBar}{
  \vspace{0.2cm}
  \includegraphics[width=16cm]{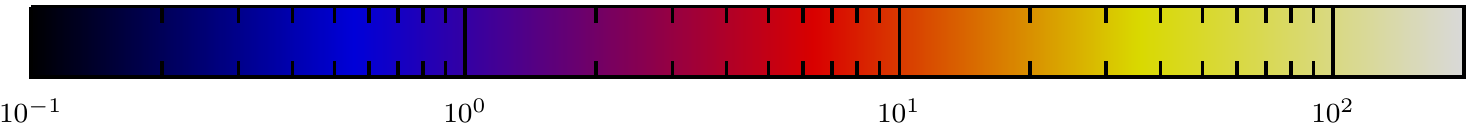}\\
  column density [g~cm$^{-2}$]\\\vspace{0.2cm}
}

\newcommand{\rhoav}{\langle\rho\rangle}
\newcommand{\rkl}[1]{\left(#1\right)}
\newcommand{\ekl}[1]{\left[#1\right]}

\newcommand{\skl}[1]{\left\langle#1\right\rangle}

\newcommand{\EkinOverEpotf}{\frac{E_\text{kin}}{|E_\text{pot}|}}
\newcommand{\EthermOverEpotf}{\frac{E_\text{therm}}{|E_\text{pot}|}}

\begin{document}

\maketitle
\begin{abstract}

We present a detailed parameter study of collapsing turbulent cloud cores, varying the initial density profile and the initial turbulent velocity field. We systematically investigate the influence of different initial conditions on the star formation process, mainly focusing on the fragmentation, the number of formed stars, and the resulting mass distributions. Our study compares four different density profiles (uniform, Bonnor-Ebert type, $\rho\propto r^{-1.5}$, and $\rho\propto r^{-2}$), combined with six different supersonic turbulent velocity fields (compressive, mixed, and solenoidal, initialised with two different random seeds each) in three-dimensional simulations using the adaptive-mesh refinement, hydrodynamics code FLASH. The simulations show that density profiles with flat cores produce hundreds of low-mass stars, either distributed throughout the entire cloud or found in subclusters, depending on the initial turbulence. Concentrated density profiles always lead to the formation of one high-mass star in the centre of the cloud and, if at all, low-mass stars surrounding the central one. In uniform and Bonnor-Ebert type density distributions, compressive initial turbulence leads to local collapse about 25\% earlier than solenoidal turbulence. However, central collapse in the steep power-law profiles is too fast for the turbulence to have any significant influence. We conclude that (I) the initial density profile and turbulence mainly determine the cloud evolution and the formation of clusters, (II) the initial mass function (IMF) is not universal for all setups, and (III) that massive stars are much less likely to form in flat density distributions. The IMFs obtained in the uniform and Bonnor-Ebert type density profiles are more consistent with the observed IMF, but shifted to lower masses.
\end{abstract}


\section{Introduction}
The current paradigm of present-day star formation suggests that stars are born in molecular clouds, permeated by supersonic turbulence (\citealt{Elmegreen04}, \citealt{MacLow04}, \citealt{Ballesteros07}). The cores have sizes of a few tenths of a parsec, are very dense with $\skl{n}\sim10^6~\text{cm}^{-3}$ \citep{Beuther07}, and in many cases they show large line widths, indicating supersonic, turbulent motions with a power-law spectral velocity distribution consistent with $P(k)\!\propto\!k^{-2}$ \citep{Zuckerman74,Larson81,Heyer04}, and thus steeper than the Kolmogorov spectrum of turbulence, $P(k)\!\propto\!k^{-5/3}$. The steeper power-law exponent is a result of the compressible cascade of interstellar turbulence \citep{Federrath10b}, in contrast to the incompressible cascade in Kolmogorov turbulence. The star-forming regions are observed to be fragmented with a filamentary, fractal-like structure \citep[][and reference therein]{Scalo1990,Menshchikov10}. Very dense cores that are supposed to form massive stars have higher temperatures ($T\sim20~\text{K}$) in contrast to less dense clouds with $10~\text{K}$ (\citealt{Beuther07}, \citealt{Ward-Thompson07}).

Despite different fragmentation structures and different local environments, the overall interplay of physical processes that contribute to the formation of stars seems to be very robust in producing prestellar cores and finally stars with a mass distribution that does not show significant differences in most observed regions of our local Universe. This mass distribution can be described by a universal initial mass function (IMF) \citep{Scalo86, Scalo98, Kroupa01, Chabrier03}. Only under extreme circumstances, i.e. close to the Galactic Centre, may the initial mass function differ from the universal one. Whereas \citet{Loeckmann10} find that even there star formation is consistent with the canonical IMF, \citet{Bartko10} clearly exclude a standard IMF in favour of a top-heavy mass function in the Galactic Centre stellar disks.

We know from observations that star formation is a complex interplay between a number of physical processes and ingredients: gravity, turbulence, rotation, radiation, thermodynamics, and magnetic fields. However, to what extent the various processes have a dominant impact on the evolution in comparison to the initial conditions of the molecular cloud is still unclear. Especially the impact of the initial conditions on the formation of massive stars, the spatial distribution of stars, and the mass evolution is unknown. Observations reveal that massive stars form early and with a tendency to be located at the centre of the cloud, whereas stars with lower masses form further out and at later times (\citealt{Hillenbrand97}; \citealt{Hillenbrand98}; \citealt{Fischer98}; \citealt{deGrijs02}; \citealt{Sirianni02}; \citealt{Gouliermis04}; \citealt{Stolte06}; \citealt{Sabbi08}). 

Theoretical approaches reproduce consistent star formation key data with a variety of different numerical methods, initial setups, and physical processes \citep[see review by][]{Klessen09}. However, a systematic study of how the initial conditions influence the fragmentation process, the collapse of the gas into stars, the number of stars, and their accretion history is still missing. Especially how the formation of massive protostars depends on the interplay between initial density profile, turbulence, and accretion model needs to be studied systematically. The large variety of existing numerical simulations all with different initial conditions does not allow for a useful comparison. \citet{Bate03}, \citet{Bate05}, \citet{Bate09,Bate09a,Bate09b}, \citet{Clark08a}, \citet{Bonnell03,Bonnell04}, and \citet{Bonnell05} used uniform density distributions with solenoidal (divergence-free), decaying turbulent motions on different cloud scales. They use a turbulent power spectrum, $P(k)\propto k^{-2}$, consistent with supersonic turbulence, however, the influence of different mixtures of initial modes of the turbulence were never investigated. In particular, \citet{Bate09a} concluded from the similarity of their results with two different initial turbulence spectra, $P(k)\!\propto\!k^{-2}$ versus $P(k)\!\propto\!k^{-3}$, that different turbulence in general has no major influence on star formation. However, both of the investigated spectra in \citet{Bate09a} are steep, such that the turbulence is dominated by the few large-scale modes (low $k$) anyway. Different mixtures of solenoidal and compressive modes of the initial turbulence are expected to have a much stronger influence on star formation, which we show here. \citet{Krumholz07,Krumholz10} favour concentrated density profiles with $\rho\propto r^{-1.5}$, referring to observations of dense cores. Their decaying turbulent velocities are based on a power spectrum of the form $P(k)\propto k^{-2}$, but not specifying the nature of the modes. In contrast, \citet{Klessen01} used driven turbulence on different scales to create dense cores self-consistently with a $\rho\propto r^{-2}$ density profile in the outer region. \citet{Offner08} compared driven and undriven turbulence with an initial flat power spectrum for the wave numbers $3\le k\le 4$. \citet{Federrath08,Federrath09,Federrath10b} investigated purely driven turbulence with the two limiting mixtures of turbulent modes: 1) fully solenoidal (divergence-free) and 2) fully compressive (curl-free), and found significantly different density distributions, with three times larger standard deviations of the density probability distribution function in the case of compressive compared to solenoidal driving \citep[see also the follow-up studies by][]{SchmidtEtAl2009,SchmidtEtAl2010,SeifriedEtAl2010,PriceFederrathBrunt2010}. Since such strongly different density fields are expected to lead to very different modes of star formation, we also investigate here three mixtures of the initial turbulence (compressive, mixed, and solenoidal). Here, however, we only apply the different turbulent modes as an initial condition, not continuously replenishing them by driving.

In this work, we combine four different extreme density profiles with different turbulent velocity fields to study the influence of the initial conditions on the formation of stars. The mass of the cloud is kept constant for all simulations. We investigate the fragmentation, the time scales, and the stellar distributions with a focus on how different initial conditions lead to different morphology and statistics of prestellar cores and stellar clusters.

The paper is structured as follows: section~\ref{sec:numerics_initcond} describes the initial density profiles and the applied turbulent velocity fields for the simulations, as well as the numerical key parameters, and the usage of sink particles. In addition, a theoretical estimate of the accretion rate for the $\rho\!\propto\!r^{-2}$ density profiles is calculated. In section~\ref{sec:results} we present the results of the simulations, followed by a discussion in section~\ref{sec:discussion}. Here we concentrate on the cloud evolution and the global stellar properties. A detailed investigation of the spatial stellar distribution will be published in a separate paper. Finally, in section~\ref{sec:summary_conclusion} we summarise our results and conclusions.

\section{Numerical Methods \& Initial Conditions}
\label{sec:numerics_initcond}
\subsection{Global Simulation Parameters}
We simulate the collapse of an initially spherical molecular cloud with a radius of $R_0=3\times10^{17}~\text{cm} \approx 0.097$~pc, centred in a cubic computational domain of length $L_\text{box}=8\times10^{17}~\text{cm}$. The gas with a mean molecular weight of $\mu = 2.3$ is assumed to be isothermal at a temperature of $20$~K. The isothermal sound speed is given by
\begin{equation}
  c_\text{s}=\sqrt{\frac{k_\text{B}T}{\mu m_\text{p}}\,} = 0.268~\text{km s}^{-1}
\end{equation}
with the Boltzmann constant $k_\text{B}$, the temperature $T$, the molecular weight $\mu$, and the proton mass $m_\text{p}$. For all runs the total mass enclosed within this sphere is $100~M_\odot$. The resulting average density is $\rhoav = 1.76\times10^{-18}~\text{g~cm}^{-3}$ or $\langle n \rangle = 4.60\times10^5~\text{cm}^{-3}$, leading to a free-fall time
\begin{equation}
\label{eq:free-fall-time}
  t_\text{ff} = \sqrt{\frac{3\pi}{32\,G\,\rhoav}\,}
\end{equation}
of $1.58\times10^{12}$~s or $50.2$~kyr. However this global average time is not a good measure for the strongly concentrated density profiles, where star formation and gravitational collapse occurs on much shorter time scales.
All of the initial spheres are gravitationally highly unstable. With the Jeans length
\begin{equation}
  \label{eq:jeans-length}
  \lambda_\text{J} = \sqrt{\frac{\pi c_\text{s}^2}{G\rhoav}\,} = 9264~\text{AU} = 0.46~R_0
\end{equation}
the Jeans volume, given as a sphere with diameter $\lambda_\text{J}$, reads $V_\text{J} = \pi\lambda_\text{J}^3/6$ and the Jeans mass of this sphere is $M_\text{J} = V_\text{J}\,\rhoav = 1.23~M_\odot$.
The central region inside the Jeans volume is called the `core' in the following.
Accounting for the different masses inside the Jeans core due to different central mass concentrations $M(r=\lambda_\text{J}/2) = M^\text{core}$, it is useful to define the new average density ($\rho^\text{core}$) and free-fall time ($t_\text{ff}^\text{core}$) for the core region $V_J$. An overview of all the physical parameters is given in table~\ref{tab:phys-param}, the core values for the different density profiles can be seen in table~\ref{tab:instability}.
\begin{table}
  \caption{Physical parameters}
  \label{tab:phys-param}
  \begin{tabular}{lcc}
    Parameter & & Value\\
    \hline
    cloud radius & $R_0$ & $3\times10^{17}\,\mathrm{cm}\approx0.097\,\mathrm{pc}$\\
    total cloud mass & $M_\text{tot}$ &$100~M_\odot$\\
    mean mass density & $\rhoav$ & $1.76\times10^{-18}$~g~cm$^{-3}$\\
    mean number density & $\langle n \rangle$ & $4.60\times10^5$~cm$^{-3}$\\
    mean molecular weight & $\mu$ & $2.3$\\
    temperature & $T$ & 20~K\\
    sound speed & $c_\text{s}$ & $2.68\times10^4$~cm~s$^{-1}$\\
    rms Mach number & $\mathcal{M}$ & $3.28-3.64$\\
    mean free-fall time & $t_\text{ff}$ & $5.02\times10^4~\text{yr}$\\
    sound crossing time & $t_\text{sc}$ & $7.10\times10^5~\text{yr}$\\
    turbulent crossing time & $t_\text{tc}$ & $1.95 - 2.16\times10^5~\text{yr}$\\
    Jeans length & $\lambda_\text{J}$ & $9.26\times10^3\,\mathrm{AU}\approx0.23~R_0$\\
    Jeans volume & $V_\text{J}$ & $1.39\times10^{51}~\text{cm}^3$\\
    Jeans mass & $M_\text{J}$ & $1.23~M_\odot$\\
    \hline
  \end{tabular}
  
  \medskip
  List of the physical parameters of the runs, which are the same for all setups.
\end{table}

\begin{table*}
  \begin{minipage}{126mm}
    \caption{Core properties of the different density distributions}
    \label{tab:instability}
    \begin{tabular}{lcccccc}
      setup & $M^\text{core}$ [$M_\odot$] & $\rho^\text{core}$ [g~cm$^{-3}$] & $n^\text{core}$ [cm$^{-3}$]& $t_\text{ff}^\text{core}$ [kyr] & $t_\text{tc}^\text{core}/t_\text{ff}^\text{core}$ \\
      \hline
      TH   &  $\phantom{0}1.25$ & $1.76\times10^{-18}$ & $4.60\times10^{5}$ & 49.858 & 1.64\\
      BE   &  $\phantom{0}5.84$ & $8.33\times10^{-18}$ & $2.18\times10^{6}$ & 23.061 & 2.12\\
      PL15 & $11.12$ & $1.59\times10^{-17}$ & $4.16\times10^{6}$ & 16.707 & 2.92\\
      PL20 & $23.02$ & $3.29\times10^{-17}$ & $8.61\times10^{6}$ & 11.615 & 4.20\\
      \hline
    \end{tabular}
    
    \medskip
    Core masses, densities, and free-fall times inside a sphere with diameter of a Jeans length ($r^\text{core} = \lambda_\text{J}/2 = 7\times10^{16}~\text{cm}$). The free-fall time for the top-hat differs slightly from the theoretical value calculated by equation~(\ref{eq:free-fall-time}), because the data from this table are the numerical values taken from the simulation.
  \end{minipage}
\end{table*}

The simulations do not include radiative feedback nor magnetic fields. The simulated density range justifies an isothermal equation of state. However, the missing heating effect due to radiation leads to more collapsing regions than in non-isothermal simulations. We therefore over-estimate the number of formed protostars, and the presented stellar statistics should more be understood as a comparison between the runs rather than an exact measurement of the IMF. 

\subsection{Numerical Code}
The simulations were carried out using the astrophysical code FLASH \citep{FLASH00} in version 2.5 which integrates the hydrodynamic equations with a piecewise-parabolic method (PPM) \citep{Colella84}. The code is parallelised using MPI. The computational domain is subdivided into blocks containing a fixed number of cells with an adaptive mesh refinement (AMR) technique based on the PARAMESH library \citep{PARAMESH99}.

\subsection{Resolution and Sink Particles}
For the main simulations an effective resolution of $4096^3$ cells was used, corresponding to a smallest cell size of $\Delta x\approx 13$~AU. In order to avoid artificial fragmentation, the Jeans length
\begin{equation}
    \lambda_\text{J} = \sqrt{\frac{\pi c_\text{s}^2}{G\rho_\text{max}}\,}
\end{equation}
at this effective resolution has to be resolved with at least $4$ grid cells \citep{Truelove97}. With sink particles, the accretion radius has to be at least $2$ grid cells at the highest level of refinement in order to fulfil this criterion. In our simulations we use an accretion radius of $3\,\Delta x$, leading to a threshold density $\rho_\text{max}$ of
\begin{equation}
  \rho_\text{max} = \frac{\pi c_\text{s}^2}{4\,G\,(3\,\Delta x)^2} = 2.46\times10^{-14}\text{g~cm}^{-3}.
\end{equation}
As heating of molecular gas begins at a density of about $10^{-13}~\text{g~cm}^{-3}$, the assumption of an isothermal equation of state seems justified \citep[e.g.,][]{Larson69}. However, we will see in the results section that fragmentation is slightly overestimated with the assumption of an isothermal equation of state up to these densities \citep[see also,][]{Krumholz07,Bate09b}.

We apply the sink particle creation criteria of \citet{Federrath10a} to avoid transient density fluctuations to be erroneously turned into sink particles, and thus to avoid artificial fragmentation. If the density in a cell on the highest level of the adaptive mesh hierarchy exceeds the resolution limit, $\rho_\text{max}$, a spherical control volume with a radius of 3 cells at the highest level of refinement ($r_\text{accr}\approx39$~AU) around that cell is investigated for collapse indicators. An accreting Lagrangian sink particle is only formed if the gas in this control volume:
\begin{itemize}
\item is converging along all principal axis, $x$, $y$, and $z$,
\item has a central minimum of the gravitational potential,
\item is Jeans-unstable,
\item is gravitationally bound,
\item is not within $r_\text{accr}$ of an existing sink particle.
\end{itemize}
The numerical parameters for the sink particles are listed in table~\ref{tab:simul-param}.

\begin{table}
  \caption{Numerical simulation parameters}
  \label{tab:simul-param}
  \begin{tabular}{lcc}
    Parameter & & Value\\
    \hline
    simulation box size & $L_\text{box}$ & $8\times10^{17}$~cm\\
    smallest cell size & $\Delta x$ & $13.06$~AU\\
    Jeans length resolution & & $\ge6$ cells\\
    max. gas density & $\rho_\text{max}$ & $2.46\times10^{-14}$~g~cm$^{-3}$\\
    max. number density& $n_\mathrm{max}$ & $6.45\times10^{9}$~cm$^{-3}$\\
    sink particle accretion radius &$r_\text{accr}$ & $39.17$~AU\\
    \hline
  \end{tabular}

  \medskip
  General simulation parameters that are related to the numerical resolution.
\end{table}

\subsection{Initial Density Profiles}

In the simulations the following four frequently used initial density profiles are applied:
\begin{enumerate}
\item Uniform density profile (Top-hat, TH)
\item Rescaled Bonnor-Ebert sphere (BE)
\item Power-law profile $\rho\propto r^{-1.5}$ (PL15)
\item Power-law profile $\rho\propto r^{-2.0}$ (PL20).
\end{enumerate}

The profiles are motivated by the following reasonings. The TH just reflects the initial conditions in a uniform density environment with finite size. Neither initial density perturbations have been established nor does the sphere have a developed over-density. The BE profile is motivated by the theoretical calculation of an isothermal sphere in hydrostatic equilibrium confined by external pressure (\citealt{Ebert55}, \citealt{Bonnor56}). The PL20 profile is the limit of the collapsing BE sphere at the end of the evolution process. This density configuration of a singular isothermal sphere is widely applied because its collapse can be described by a self-similar solution with predictable in-fall and evolution properties (\citet{Shu77}, section~\ref{sec:PL20-self-similarity}). So far studies with a singular isothermal sphere have only been done without turbulent velocity. Finally the PL15 profile, which is an intermediate evolutionary stage of the BE sphere before reaching the PL20 configuration, is motivated by observations. The outer region of collapsing clouds is observed to follow a density distribution of the form $\rho\propto r^{-1.6}$ \citep{Pirogov09}.

\begin{figure}
  \centering
  \includegraphics[height=8cm]{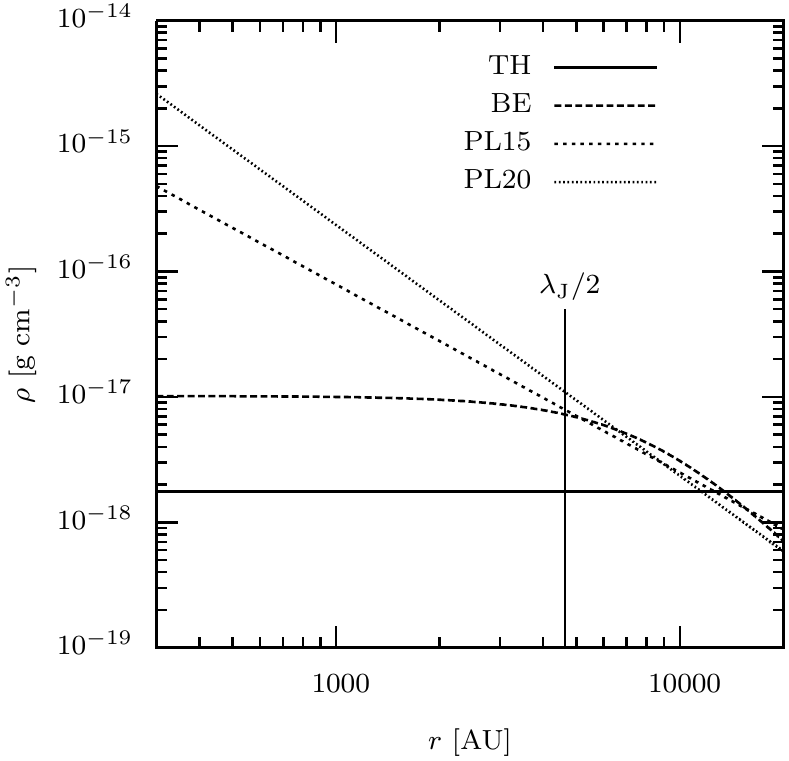}
  \caption{Comparison of the four initial density profiles adjusted to a total mass of $100~M_\odot$ within a radius of $0.1$~pc. $\lambda_\text{J}$ marks the Jeans length for the average density $\rhoav$.}
  \label{fig:dens-profile-comparison}
\end{figure}

A comparison of the radial shape for all density profiles is shown in figure~\ref{fig:dens-profile-comparison}. $\lambda_\text{J}$ marks the Jeans length for the average density $\rhoav$. These four profiles are extreme setups that allow us to follow the influence on the central collapse and the fragmentation.

No initial density fluctuations were applied. The density of the surrounding gas in the cubic box around the spherical molecular cloud is set to $10^{-2}$ times the gas density at the edge of the cloud at $r=R_0$. The initial temperature distribution is a step function with the temperature in the cloud envelope $100$ times larger than in the inner isothermal collapsing cloud, which results in a continuous pressure at the boundary $r=R_0$.

\subsubsection{Top-hat}
This density implementation is the simplest profile, describing the gas density as a step function
\begin{equation}
  \rho =
  \begin{cases}
    \rhoav\quad&\text{for}~r \le R_0,\\
    0.01~\rhoav\quad&\text{for}~r > R_0\\
  \end{cases}
\end{equation}
with
\begin{equation}
  \rhoav = \frac{M_\text{tot}}{V} = \frac{3 M_\text{tot}}{4\pi R_0^3}.
\end{equation}

\subsubsection{Rescaled Bonnor-Ebert Sphere}
In hydrostatic equilibrium the critical density profile is described by a Bonnor-Ebert sphere with normalised radius $\xi=6.41$ (\citealt{Ebert55}, \citealt{Bonnor56}). The only free parameter for this configuration is the central density $\rho_0$. In order to better compare this sphere with the other clouds, the central density was first chosen such that the outer radius of the sphere yielded the given size of $0.1$~pc. Then the density at every point was rescaled to fit the total cloud mass of $M_\text{tot}=100~M_\odot$.

\subsubsection{Power-law Profiles}
\label{sec:PLs}
As the power-law profiles $\rho\propto r^{-p}$ diverge in the centre of the cloud, an inner radius has to be defined below which the density follows a finite function. In these setups this part of the profile is described by a quadratic function:
\begin{equation}
  \rho =
  \begin{cases}
    ar^2+c \quad&\text{for }0\le r < r_1,\\
    B\,\left(\frac{r}{R_0}\right)^{-p} \quad&\text{for }r_1\le r \le R_0.
  \end{cases}
\end{equation}
The reason for this transition instead of a simple cut-off at the inner radius is to avoid artificial numerical effects at the boundary $r_1$. The value for $r_1$ was set to $3$ ($5$) times the cell size at the highest level of refinement for $p=1.5$ ($p=2.0$). The choice for the values of $a$ and $c$ allow for a continuous transition for the density function value as well as for the derivative $\text{d}\rho/\text{d}r$. For $p=1.5$ the two values read $a=2.227\times10^{-44}~\text{g cm}^{-5}$ and $c=1.784\times10^{-14}~\text{g cm}^{-3}$, the values for $p=2.0$ are $a=5.804\times10^{-44}~\text{g cm}^{-5}$ and $c=1.107\times10^{-13}~\text{g cm}^{-3}$.
The outer radius $R_0$ was set to the radius of the cloud, the constant $B$ scales the density profile to a total enclosed mass of $M_\text{tot} = 100~M_\odot$. Its value depends on the inner radius $r_1$. However, for small radii $r_1$, which is roughly three orders of magnitude smaller than $R_0$ in the numerical setup, $B$ converges to
\begin{equation}
  \lim_{r_1\rightarrow 0}B=\frac{M_\text{tot}(3-p)}{4\pi}\,\frac{1}{R_0^3}.
\end{equation}
Depending on the effective resolution and therefore the parameter $r_1$, the maximum density changes significantly.

\subsubsection{Power-law Profile $\rho\propto r^{-2}$ and Self-similarity}
\label{sec:PL20-self-similarity}
Based on the analytic treatment of the collapse of a singular isothermal sphere by \citet{Shu77}, the evolution of a density profile with the general form
\begin{equation}
  \rho(r, t>0) = \frac{c_\text{s}^2}{2\pi G}r^{-2},\qquad c_\text{s}^2 = \frac{k_\text{B}T}{\mu m_\text{p}}
\end{equation}
can be described using the dimensionless similarity variable
\begin{equation}
  x = \frac{r}{c_\text{s}t},
\end{equation}
where $G$ is the gravitational constant. The density distribution, the mass accretion rate, and the in-fall velocity can be transformed to
\begin{align}
  \rho(r,t) &= \frac{\alpha(x)}{4\pi G t^2}\\
  \dot{M}_\text{SIS}(r,t) &= \frac{c_\text{s}^3}{G}\,m(x)\\
  u(r, t) &= c_\text{s}\, v(x)
\end{align}
with $\alpha(x) = x^{-2}~\text{d}m/\text{d}x$ such that the collapse proceeds in a self-similar way. The two basic differential equations that have to be solved in order to find the values for $\alpha$ and $v$ read
\begin{align}
  \label{eq:Shu-ODE}
  \ekl{\rkl{x-v}^2-1}\frac{\text{d}v}{\text{d}x} &= \ekl{\alpha\rkl{x-v}-\frac 2 x}\rkl{x-v}\\
  \ekl{\rkl{x-v}^2-1}\frac 1 \alpha \frac{\text{d}\alpha}{\text{d}x} &= \ekl{\alpha-\frac 2 x \rkl{x-v}}\rkl{x-v}\notag.
\end{align}
The initial density profile must have the form
\begin{equation}
  \label{eq:Shu-general-density}
  \rho(r, t=0) = \frac{c_\text{s}^2\,A}{4\pi G}\,r^{-2}
\end{equation}
with $A>2$. This equation can be rewritten for the PL20 density setup as
\begin{equation}
  \rho(r,t=0) = q\,r^{-2}\quad\text{with}\quad q=5.30\times10^{16}~\text{g~cm}^{-1}
\end{equation}
for a total enclosed mass of $100~M_\odot$.
The constant $A$ in this setup has the value
\begin{equation}
  A = \frac{4\pi Gq}{c_\text{s}^2} \approx 61.9.
\end{equation}
Comparing the factor $A$ to the number of Jeans masses in the cloud
\begin{equation}
  M_\text{J} = \frac{\pi^{5/2}}{6}\frac{c_\text{s}^3}{G^{3/2}\rho^{1/2}},
\end{equation}
\begin{equation}
  N_\text{J} = \frac{M_\text{tot}}{M_\text{J}}
\end{equation}
it can be rewritten as follows to
\begin{equation}
  A = \frac{4\pi^{8/3}q}{6^{2/3}}\frac{N_\text{J}^{2/3}}{\rho^{1/3}M_\text{tot}^{2/3}}\propto N_\text{J}^{2/3}.
\end{equation}

\begin{figure*}
  \begin{minipage}{\textwidth}
    \includegraphics[height=8cm]{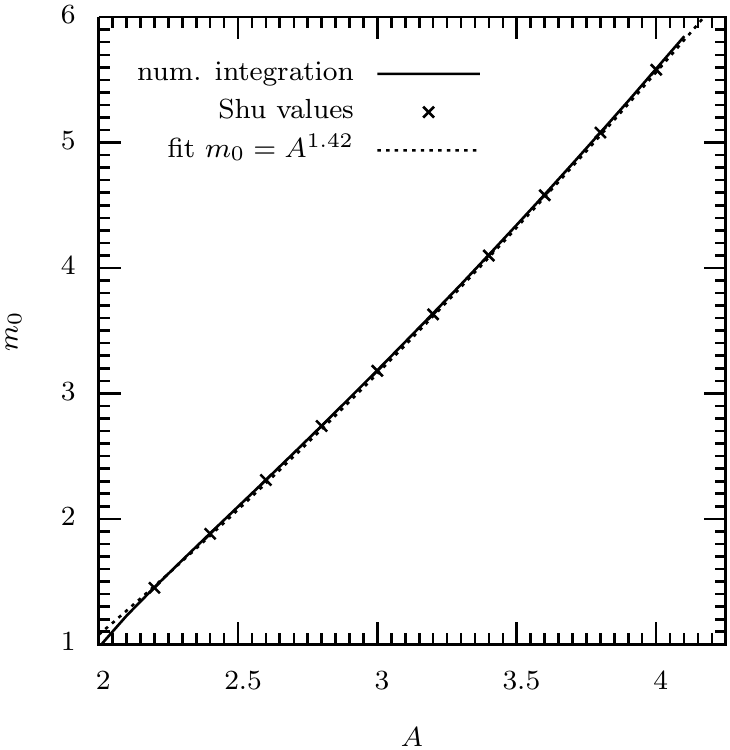}
    \includegraphics[height=8cm]{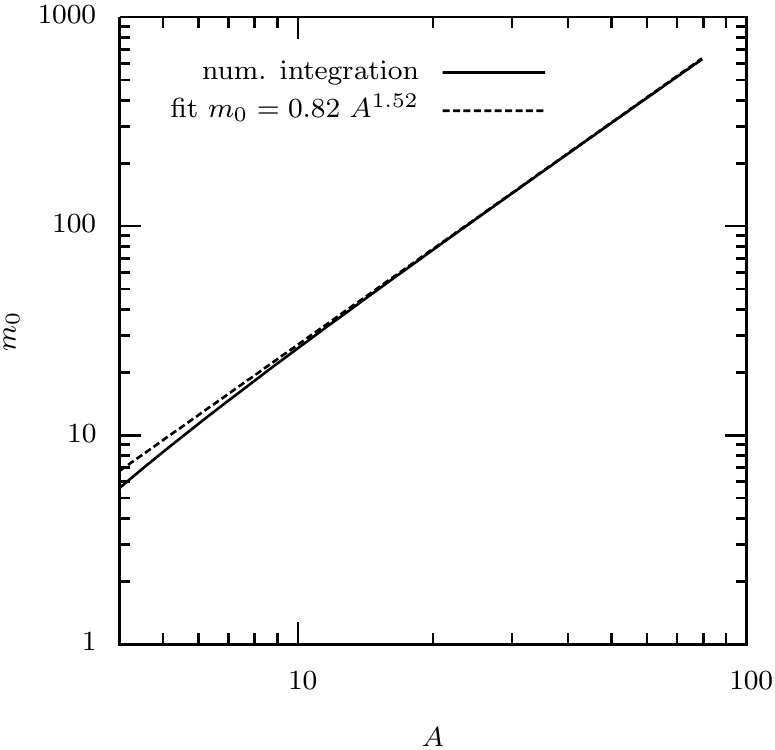}
    \caption{Accretion rates as a function of $A$ from equations~(\ref{eq:Shu-ODE}) and (\ref{eq:Shu-general-density}). In the left plot the values for small $A$ are compared with the Shu values. The right plot shows the high-$A$ regime relevant for the simulation with the PL20 density profile.}
    \label{fig:PL20-Shu-accretion}
  \end{minipage}
\end{figure*}

In order to find the theoretical value for the accretion factor $m_0 = m(r=0,t=0)$
equations~(\ref{eq:Shu-ODE}) have to be integrated from a large $x$ to a value close to zero.
For a critical sphere with $A = 2$ this factor is $m_0=0.95$, for $A=61.9$ it reaches a very high value of $m_0\approx 421$ (see figure~\ref{fig:PL20-Shu-accretion}). This finally gives a theoretical accretion rate of
\begin{equation}
  \dot{M}_\text{SIS} = m_0\,\frac{c_\text{s}^3}{G} \approx 1.89\times10^{-3}~M_\odot~\text{yr}^{-1}.
\end{equation}
The accretion factor $m_0$ can be fitted with a power-law dependence
\begin{equation}
  m_0\propto A^{1.52}
\end{equation}
(see right plot in figure~\ref{fig:PL20-Shu-accretion}) which in turn gives a theoretical accretion rate close to a linear dependence on the number of Jeans masses
\begin{equation}
  m_0\propto N^{1.01}_\text{J}.
\end{equation}

\subsection{Initial Turbulence}
\subsubsection{Power Spectrum of the Turbulence}
The turbulence is modelled with an initial random velocity field, originally created in Fourier space, and transformed back into real space. The power spectrum of the modes is given by a power-law function in wavenumber space ($\mathbf{k}$ space) with $E_\text{k}\propto k^{-2}$, corresponding to Burgers turbulence (the value for incompressible, Kolmogorov turbulence would be $E_\text{k}\propto k^{-5/3}$ in this notation), which is consistent with the observed spectrum of interstellar turbulence \citep[e.g.,][]{Larson81,Heyer04}. The velocity field is dominated by large-scale modes due to the steep power-law exponent, $-2$, with the largest mode having the size of the simulation box. Thus, changing the slope of the power spectrum is not expected to affect the results significantly \citep[see,][]{Bate09a}. However, the random seed and the mixture of modes of the initial turbulence can potentially change the results more strongly, which we investigate in this study. Concerning the nature of the $\mathbf{k}$ modes, compressive (curl-free) are distinguished from solenoidal (divergence-free) ones. The simulation uses three types of initial fields: pure compressive fields (c), pure solenoidal (s), and a natural (random) mixture (m) of both. These choices were motivated by the strong differences found in driven turbulence simulations using purely solenoidal and purely compressive driving of the turbulence (\citealt{Federrath08,Federrath09,Federrath10b}). Note however that only decaying turbulence with compressive, mixed, and solenoidal modes are considered here. For each of these three types, two different random velocity seeds are created, leading to six different initial velocity fields in total (c-1, c-2, m-1, m-2, s-1, s-2), which are combined with the different density profiles.

No overall global rotation is imposed on the cloud. Due to the random nature of the turbulence, the net rotation, and the net angular momentum are not strictly zero. The ratio of rotational to gravitational energy is of the order of a few times $10^{-3}$.

\subsubsection{Mach numbers}
All setups have supersonic velocities. Due to different density concentrations and the resulting different refinement structure of the AMR grid, the rms velocities and their Mach number
\begin{equation}
  \mathcal{M} = \frac{v_\text{rms}}{c_\text{s}}
\end{equation}
differ slightly among the different density profiles.
Table~\ref{tab:runs} shows the Mach numbers for all the setups which vary from $\mathcal{M} = 3.28 - 3.64$ with an average of $\skl{\mathcal{M}}=3.44$.

\subsubsection{Sound Crossing Time and Turbulence Crossing Time}
\label{sec:crossing-times}
The sound crossing time through the entire sphere is
\begin{equation}
  t_\text{sc}(R_0) = 7.10\times 10^5~\text{yr},
\end{equation}
about one to two orders of magnitude higher than the core free-fall time for the TH or the PL20 profile, respectively. For the supersonic turbulence with an average gas velocity of Mach $3.44$, the average turbulence crossing time is
\begin{equation}
  t_\text{tc}(R_0) = 2.06\times 10^5~\text{yr}.
\end{equation}
The crossing times for the core region are $t_\text{sc}^\text{core} = t_\text{sc}(\lambda_\text{J}) = 1.64\times 10^5~\text{yr}$ and $t_\text{tc}^\text{core} =t_\text{tc}(\lambda_\text{J}) = 4.77\times 10^4~\text{yr}$, which is close to the global free-fall time.

\subsection{Runs}
In order to systematically investigate the influence of the initial conditions, we follow a variety of combinations of turbulence and density profiles. Table~\ref{tab:runs} gives an overview of the combinations. The BE profiles as well as the PL15 profiles are combined with all turbulent fields. As the TH runs are computationally very expensive, only the turbulent fields with mixed modes are applied. The PL20 density distribution has a very short central free-fall time and is expected to collapse and form a massive sink particle before the turbulent motions have an important impact on the cloud structure. Therefore 3 additional setups with compressive velocity field c-1 but higher rms Mach numbers (PL20-c-1b, PL20-c-1c \& PL20-c-1d) were simulated. The velocities in PL20-c-1b are twice as high as the ones in PL20-c-1; runs PL20-c-1c and PL20-c-1d have velocities 4 and 6 times as high as PL20-c-1. The rms Mach numbers are: $\mathcal{M}_\text{c-1b} = 6.57$, $\mathcal{M}_\text{c-1c} = 13.1$, $\mathcal{M}_\text{c-1d} = 19.7$ (see tab.~\ref{tab:runs}).

\begin{table*}
  \begin{minipage}{126mm}
    \caption{List of the runs and their main properties}
    \label{tab:runs}
    \begin{tabular}{cccccccccc}
      density & turbulent & seeds & name & effective & $\mathcal{M}$ & total &  total & core & core \\
      profile & modes     &       &      & resolution & & $\EkinOverEpotf$ & $\EthermOverEpotf$ & $\rkl{\EkinOverEpotf}_\text{c}$ & $\rkl{\EthermOverEpotf}_\text{c}$\\
      \hline                                 
      TH   & mix   & 1 & TH-m-1   & $4096^3$ & \phantom{0}3.3 & 0.075 & 0.047 & 0.027 & 0.038 \\
      TH   & mix   & 2 & TH-m-2   & $4096^3$ & \phantom{0}3.6 & 0.090 & 0.047 & 0.111 & 0.038 \\
      BE   & compr & 1 & BE-c-1   & $4096^3$ & \phantom{0}3.3 & 0.058 & 0.039 & 0.073 & 0.028 \\
      BE   & compr & 2 & BE-c-2   & $4096^3$ & \phantom{0}3.6 & 0.073 & 0.039 & 0.055 & 0.028 \\
      BE   & mix   & 1 & BE-m-1   & $4096^3$ & \phantom{0}3.3 & 0.053 & 0.039 & 0.018 & 0.028 \\
      BE   & mix   & 2 & BE-m-2   & $4096^3$ & \phantom{0}3.6 & 0.074 & 0.039 & 0.082 & 0.028 \\
      BE   & sol   & 1 & BE-s-1   & $4096^3$ & \phantom{0}3.3 & 0.055 & 0.039 & 0.057 & 0.028 \\
      BE   & sol   & 2 & BE-s-2   & $4096^3$ & \phantom{0}3.5 & 0.074 & 0.039 & 0.072 & 0.028 \\
      PL15 & compr & 1 & PL15-c-1 & $4096^3$ & \phantom{0}3.3 & 0.056 & 0.038 & 0.067 & 0.025 \\
      PL15 & compr & 2 & PL15-c-2 & $4096^3$ & \phantom{0}3.6 & 0.068 & 0.038 & 0.042 & 0.025 \\
      PL15 & mix   & 1 & PL15-m-1 & $4096^3$ & \phantom{0}3.3 & 0.050 & 0.038 & 0.013 & 0.025 \\
      PL15 & mix   & 2 & PL15-m-2 & $4096^3$ & \phantom{0}3.6 & 0.071 & 0.038 & 0.072 & 0.025 \\
      PL15 & sol   & 1 & PL15-s-1 & $4096^3$ & \phantom{0}3.3 & 0.053 & 0.038 & 0.052 & 0.025 \\
      PL15 & sol   & 2 & PL15-s-2 & $4096^3$ & \phantom{0}3.5 & 0.069 & 0.038 & 0.061 & 0.025 \\
      PL20 & compr & 1 & PL20-c-1 & $4096^3$ & \phantom{0}3.3 & 0.042 & 0.029 & 0.046 & 0.017 \\
      \hline                                 
      PL20 & compr & 1 & PL20-c-1b& $1024^3$ & \phantom{0}6.6& 0.170 & 0.029 & 0.192 & 0.018 \\
      PL20 & compr & 1 & PL20-c-1c& $1024^3$ & 13.1& 0.682 & 0.029 & 0.768 & 0.018 \\
      PL20 & compr & 1 & PL20-c-1d& $1024^3$ & 19.7& 1.534 & 0.029 & 1.728 & 0.018 \\
      \hline
    \end{tabular}
    
    \medskip
    In order to increase the influence of the turbulence for the PL20 profile three more runs (PL20-c-1b, PL20-c-1c, PL20-c-1d) were carried out with the same structure of the velocity field as PL20-c-1, but with rescaled absolute values by factors of 2, 4, and 6, leading to rms Mach numbers $\mathcal{M}_\text{c-1b} = 2\,\mathcal{M}_\text{c-1}$, $\mathcal{M}_\text{c-1c} = 4\,\mathcal{M}_\text{c-1}$, $\mathcal{M}_\text{c-1d} = 6\,\mathcal{M}_\text{c-1}$. See table~\ref{tab:resolution-parms} for resolution details.
  \end{minipage}
\end{table*}

\section{Results}
\label{sec:results}
\begin{figure}
  \includegraphics[width=8cm]{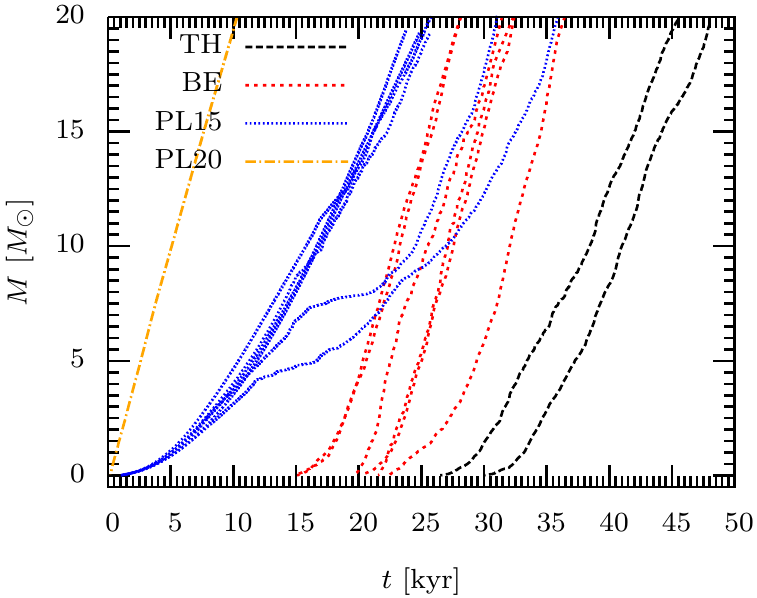}
  \caption{Comparison of the total mass in sink particles $M$ for all simulations. All velocity realisations for one density profile are combined in one line style. A detailed discussion of each velocity field is given in the analysis section of each of the density profiles.}
  \label{fig:comparison-density-profiles}
\end{figure}

We followed the collapse to a star formation efficiency of 20\%, i.e., until 20\% of the initial cloud mass was captured in sink particles. The concentrated profile PL20 reached that stage quite quickly ($\sim11~\text{kyr}$). The PL15 runs show large differences in the simulation time, ranging from $25-36~\text{kyr}$, which is similar to the time needed for the BE density setups ($27-35~\text{kyr}$). The longest time was needed for the TH setup with $45-48~\text{kyr}$. Table~\ref{tab:Nsinks_and_simtime} gives an overview of the total simulated time for all setups. Related to the core free-fall time, the TH and PL20 profiles just need roughly one $t_\text{ff}^\text{core}$ to capture $20~M_\odot$ in sink particles, whereas the BE runs need $1.2-1.5~t_\text{ff}^\text{core}$. The longest time was needed by the PL15 profiles with $1.4-2.1~t_\text{ff}^\text{core}$. A comparison of the captured mass in sink particles can be seen in figure~\ref{fig:comparison-density-profiles} for all runs. The setups with the same density profile are plotted in the same line style in order to keep the plot readable. 
\begin{table}
  \caption{Overview of the simulation time and the sink particle properties}
  \label{tab:Nsinks_and_simtime}
  \begin{tabular}{lccccc}
    Run & $t_\text{sim}$ & $t_\text{sim}/t_\text{ff}^\text{core}$ &$t_\text{sim}/t_\text{ff}$ & $N_\text{sinks}$ & $M_\text{max}$\\
    & [kyr] & & & & \\
    \hline
    TH-m-1   & 48.01 & 0.96 & 0.96 & $311$                       & $0.86$\\
    TH-m-2   & 45.46 & 0.91 & 0.91 & $429$                       & $0.74$\\
    BE-c-1   & 27.52 & 1.19 & 0.55 & $305$                       & $0.94$\\
    BE-c-2   & 27.49 & 1.19 & 0.55 & $331$                       & $0.97$\\
    BE-m-1   & 30.05 & 1.30 & 0.60 & $195$                       & $1.42$\\
    BE-m-2   & 31.94 & 1.39 & 0.64 & $302$                       & $0.54$\\
    BE-s-1   & 30.93 & 1.34 & 0.62 & $234$                       & $1.14$\\
    BE-s-2   & 35.86 & 1.55 & 0.72 & $325$                       & $0.51$\\
    PL15-c-1 & 25.67 & 1.54 & 0.51 & $194$                       & $8.89$\\
    PL15-c-2 & 25.82 & 1.55 & 0.52 & $161$                       & $12.3$\\
    PL15-m-1 & 23.77 & 1.42 & 0.48 & $\phantom{0}\phantom{0}1$   & $20.0$\\
    PL15-m-2 & 31.10 & 1.86 & 0.62 & $308$                       & $6.88$\\
    PL15-s-1 & 24.85 & 1.49 & 0.50 & $\phantom{0}\phantom{0}1$   & $20.0$\\
    PL15-s-2 & 35.96 & 2.10 & 0.72 & $422$                       & $4.50$\\
    PL20-c-1 & 10.67 & 0.92 & 0.21 & $\phantom{0}\phantom{0}1$   & $20.0$\\
    \hline                                          
    PL20-c-1b& 10.34 & 0.89 & 0.21 & $\phantom{0}\phantom{0}2$   & $20.0$\\
    PL20-c-1c& \phantom{0}9.63  & 0.83 & 0.19 & $\phantom{0}12$  & $17.9$\\
    PL20-c-1d& 11.77 & 1.01 & 0.24 & $\phantom{0}34$  & $13.3$\\
    \hline
  \end{tabular}
  
  \medskip
  The time of each simulation is given as the absolute time $t_\text{sim}$, the time in core free-fall times $t_\text{sim}/t_\text{ff}^\text{core}$, and the time in average free-fall times $t_\text{sim}/t_\text{ff}$. $N_\text{sink}$ shows the number of sink particles at the end of the run, $M_\text{max}$ gives the mass of the most massive sink particle.
\end{table}

\begin{figure*}
  \begin{minipage}{\textwidth}
    \centering
    \includegraphics[width=4.0cm]{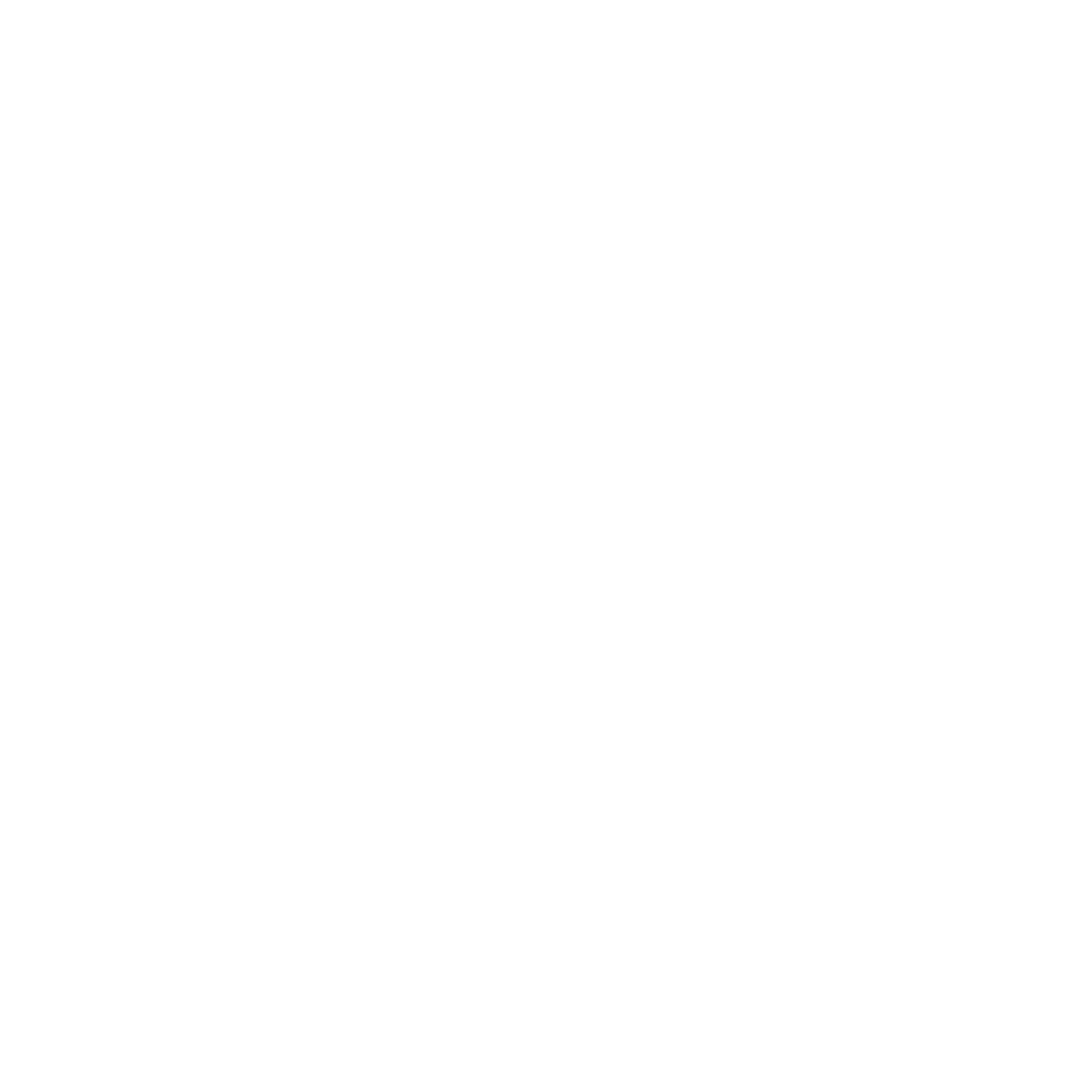}
    \includegraphics[width=4.0cm]{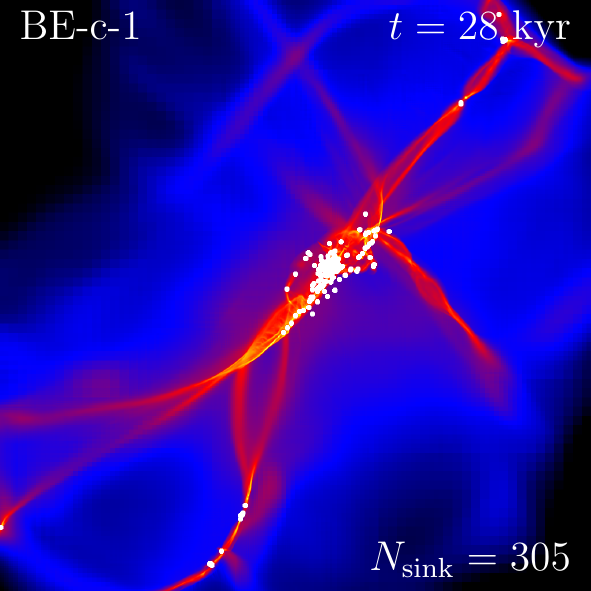}
    \includegraphics[width=4.0cm]{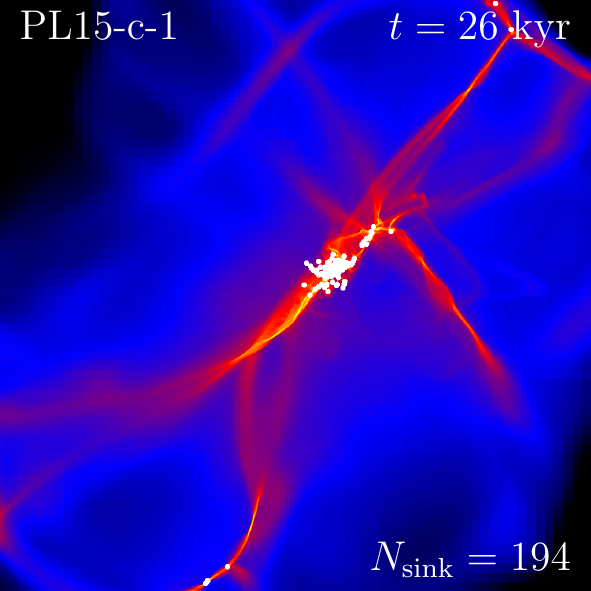}
    \includegraphics[width=4.0cm]{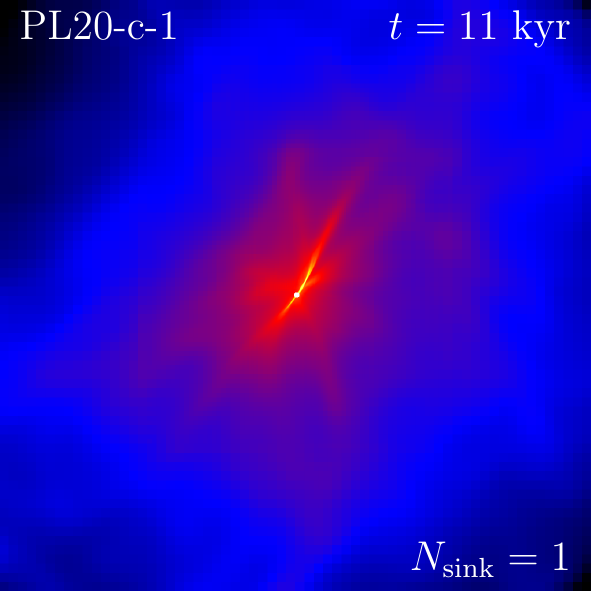}\\
    \includegraphics[width=4.0cm]{empty}
    \includegraphics[width=4.0cm]{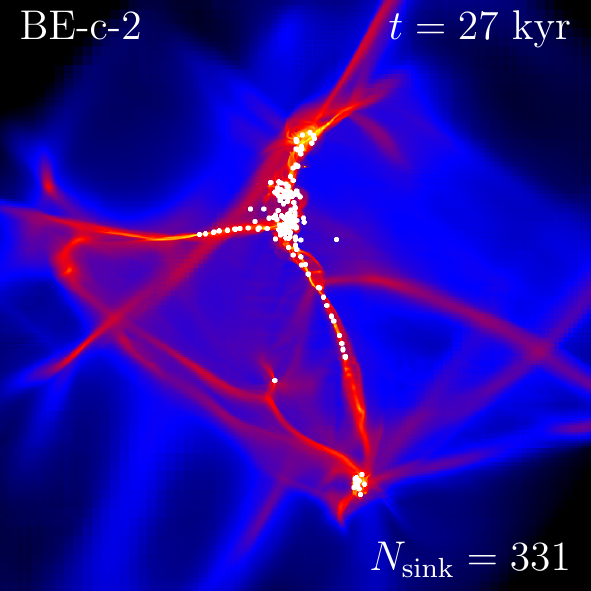}
    \includegraphics[width=4.0cm]{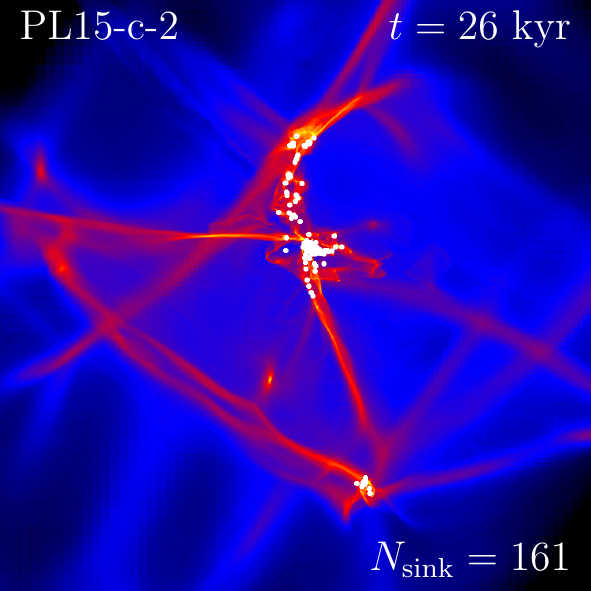}
    \includegraphics[width=4.0cm]{empty}\\
    \includegraphics[width=4.0cm]{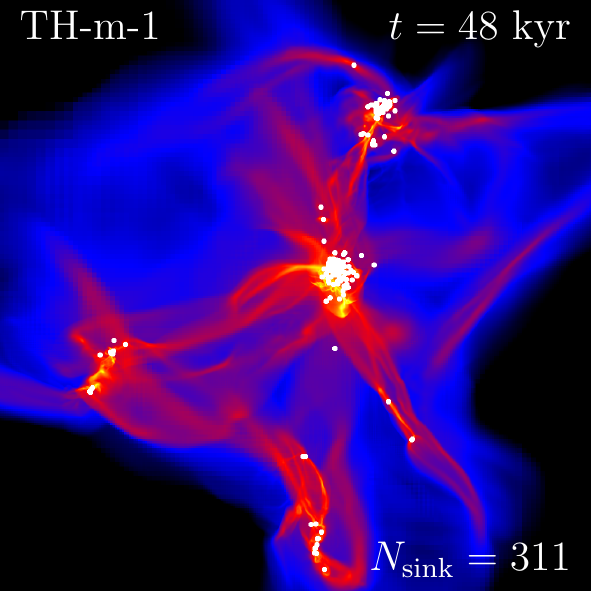}
    \includegraphics[width=4.0cm]{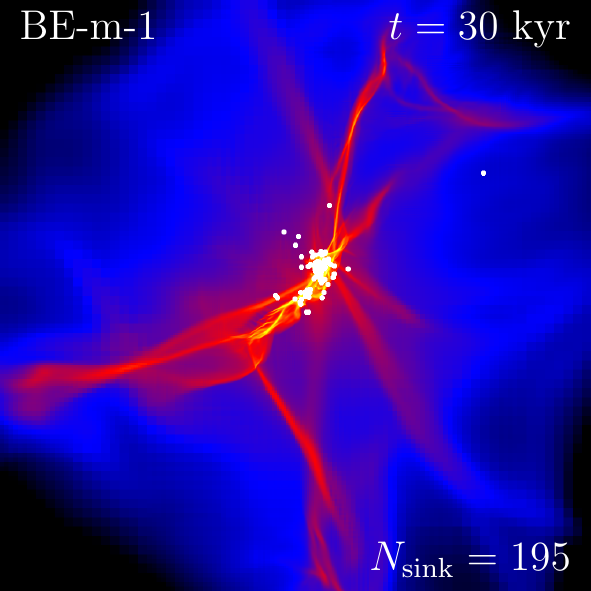}
    \includegraphics[width=4.0cm]{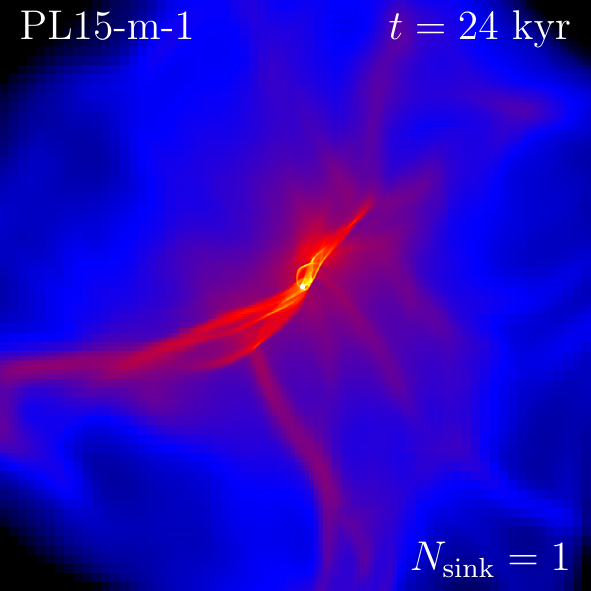}
    \includegraphics[width=4.0cm]{empty}\\
    \includegraphics[width=4.0cm]{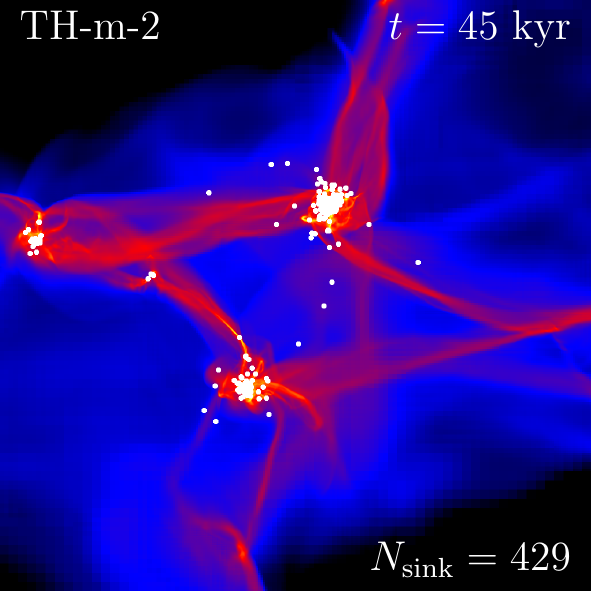}
    \includegraphics[width=4.0cm]{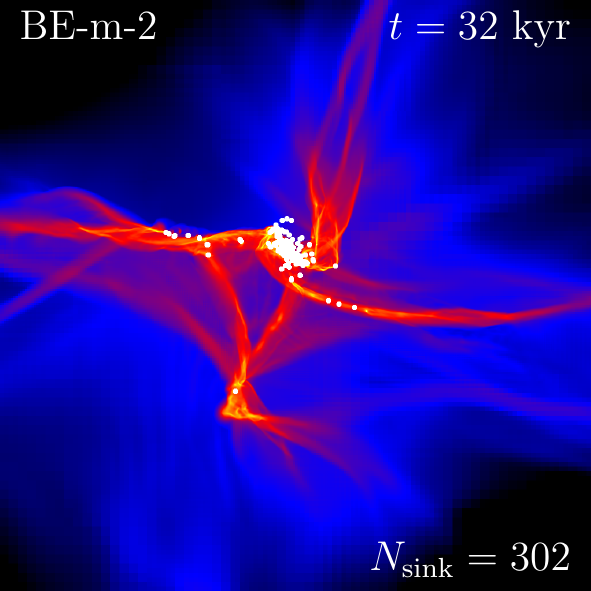}
    \includegraphics[width=4.0cm]{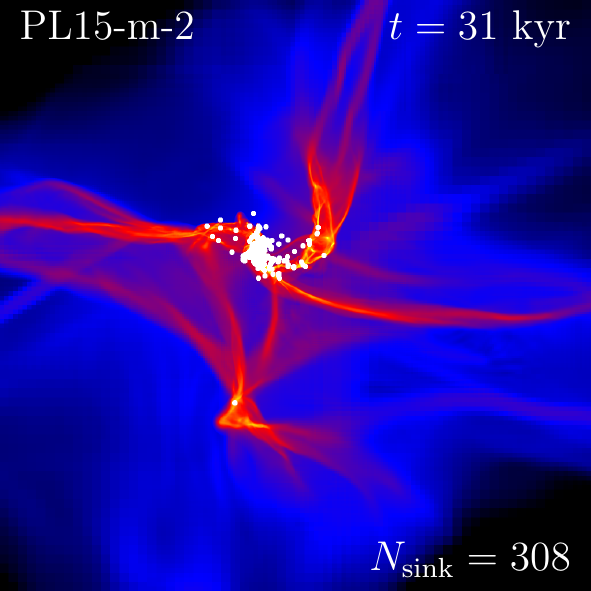}
    \includegraphics[width=4.0cm]{empty}\\
    \ColorBar
  \end{minipage}
  \caption{Column density plots for the TH, BE, and PL15 setups with velocity profiles c-1, c-2, m-1, and m-2 as well as for PL20-c-1 at the end of the simulation. The box in all cases spans $0.13$~pc in both $x$ and $y$ direction. Each picture row corresponds to one velocity field, each column to a density profile. All setups show filamentary structures but differently spread in the box. Only the TH density runs form distinct subclusters.}
  \label{fig:column-density-sim-end-part1}
\end{figure*}

\begin{figure*}
  \begin{minipage}{\textwidth}
    \centering
    \includegraphics[width=4.0cm]{empty}
    \includegraphics[width=4.0cm]{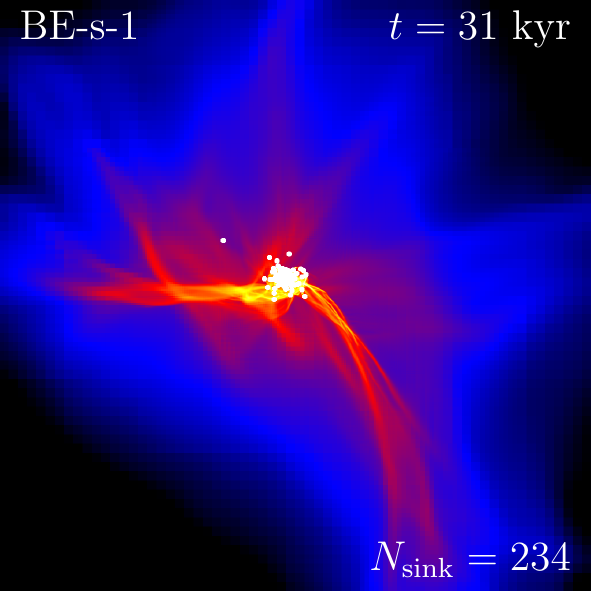}
    \includegraphics[width=4.0cm]{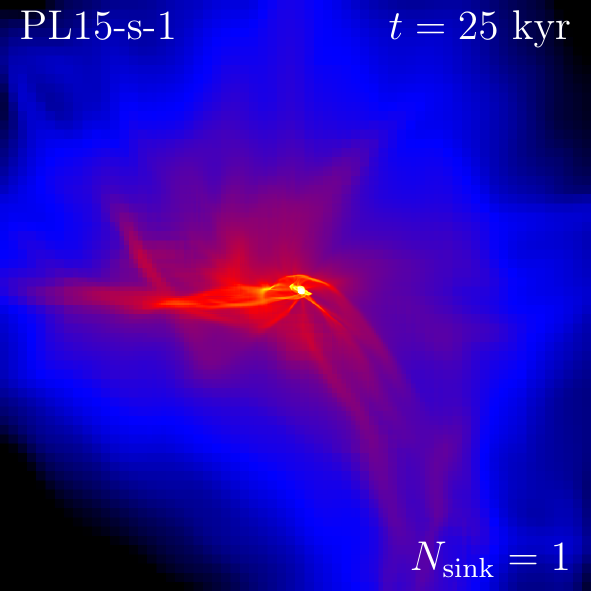}
    \includegraphics[width=4.0cm]{empty}\\
    \includegraphics[width=4.0cm]{empty}
    \includegraphics[width=4.0cm]{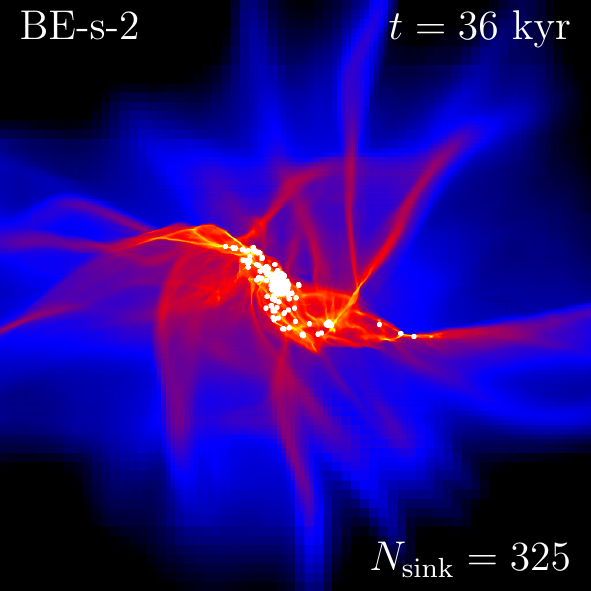}
    \includegraphics[width=4.0cm]{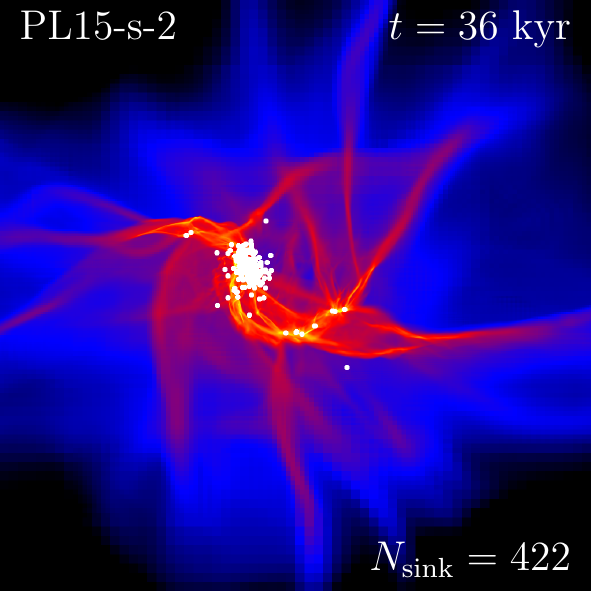}
    \includegraphics[width=4.0cm]{empty}\\
    \vspace{0.5cm}
    
    \includegraphics[width=4.0cm]{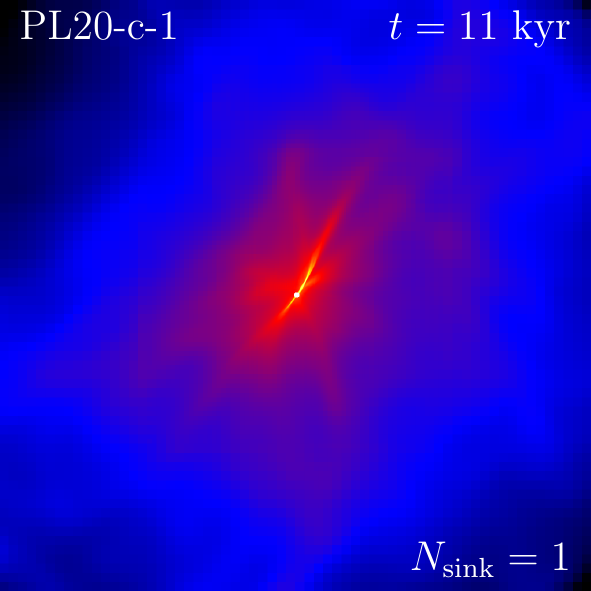}
    \includegraphics[width=4.0cm]{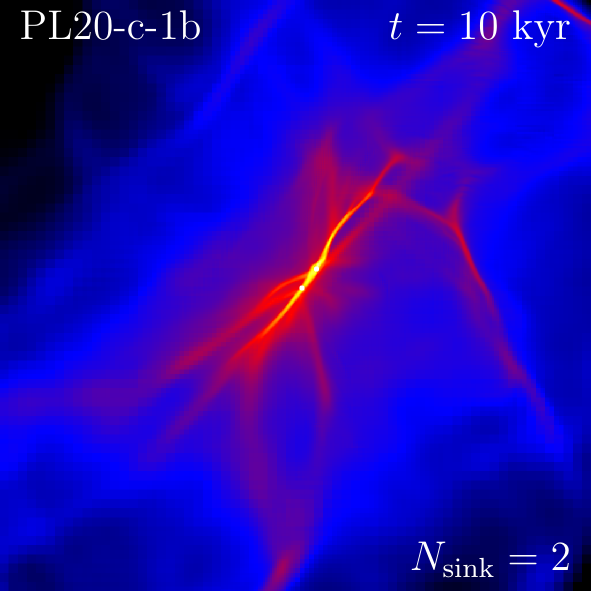}
    \includegraphics[width=4.0cm]{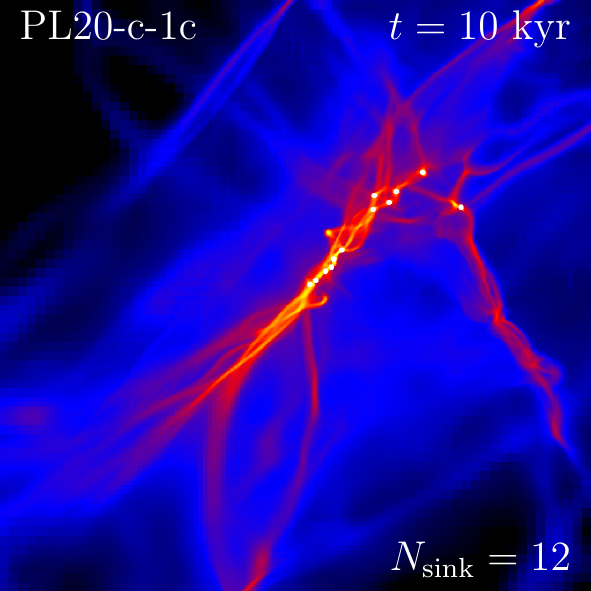}
    \includegraphics[width=4.0cm]{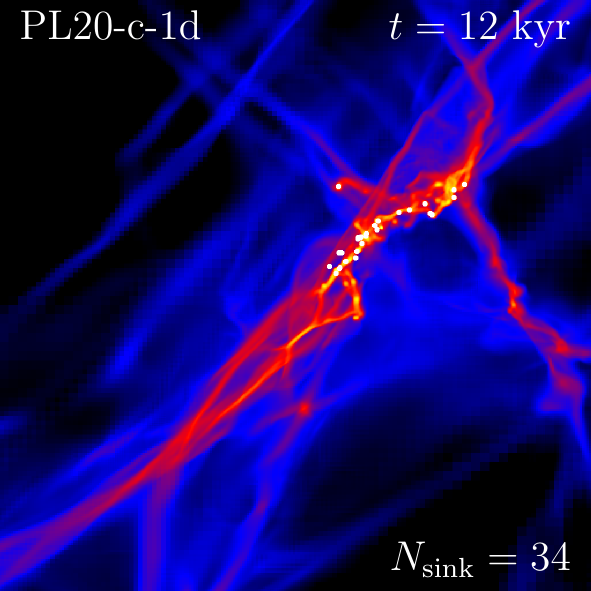}\\
    \ColorBar
  \end{minipage}
  \caption{Column density plots for the BE and PL15 setups with velocity profiles s-1 and s-2 (upper part) as well as for the PL20 setup with turbulent field c-1, c-1b, c-1c, and c-1d (lower part). The box in all cases spans $0.13$~pc in both $x$ and $y$ direction.}
  \label{fig:column-density-sim-end-part2}
\end{figure*}

During the collapse of the cloud two different gravitational processes compete with each other. Firstly, the collapse toward the centre of mass and secondly the collapse of dense regions into filaments, induced by the turbulence. The different density profiles and turbulent fields lead to different cloud evolutions, fragmentation properties, and sink particle accretion rates. A column density plot at the end of each simulation is shown in figure~\ref{fig:column-density-sim-end-part1} and \ref{fig:column-density-sim-end-part2}.
Figure~\ref{fig:column-density-sim-end-part1} shows the column density plots for the density profiles TH, BE, and PL15 with the velocity field c-1, c-2, m-1, and m-2, as well as PL20-c-1. Each picture row shows simulations with the same initial turbulent velocity field, each column belongs to one density distribution. In the upper part of figure~\ref{fig:column-density-sim-end-part2} we show the final column density for the BE and PL15 profile with the solenoidal fields. The lower part shows the PL20 profile with compressive turbulent modes for realisation 1. The four different plots belong to different initial kinetic energy variations (see table~\ref{tab:runs}). All simulations show the formation of filamentary structures and sink particles. Depending on the initial density profile, the turbulent field, and the resulting total simulation time, the position of the filaments as well as the number of sink particles and their spatial distribution vary significantly. The TH profiles in figure~\ref{fig:column-density-sim-end-part1} show locally disconnected filaments and subclusters of sink particles. The BE profiles also form many sink particles in extended filaments, but much more centrally concentrated and in stronger connected filaments. The initial mass concentration and the resulting faster central collapse suppress the formation of completely disconnected subclusters. The PL15 density profile shows in many cases a similar cloud evolution as the BE setups. However, the total number of sink particles varies strongly with different velocity realisations and the sink particles are located closer to the centre of mass. The influence of different initial kinetic energies of the turbulent motions can be seen in the PL20 setups. Higher velocities lead to much stronger substructures within the same simulated time.

\begin{figure*}
  \begin{minipage}{160mm}
    \centering
    \includegraphics[width=5.0cm]{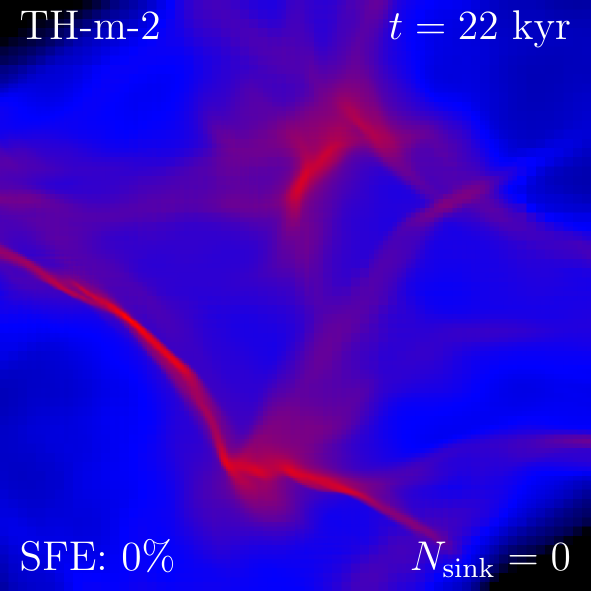}
    \includegraphics[width=5.0cm]{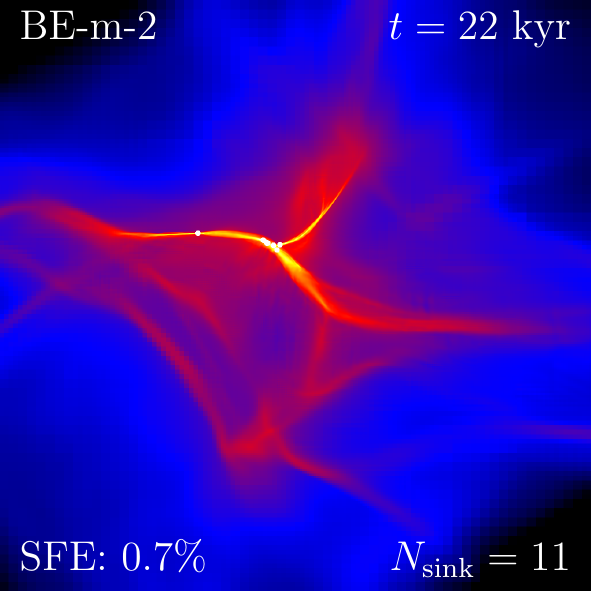}
    \includegraphics[width=5.0cm]{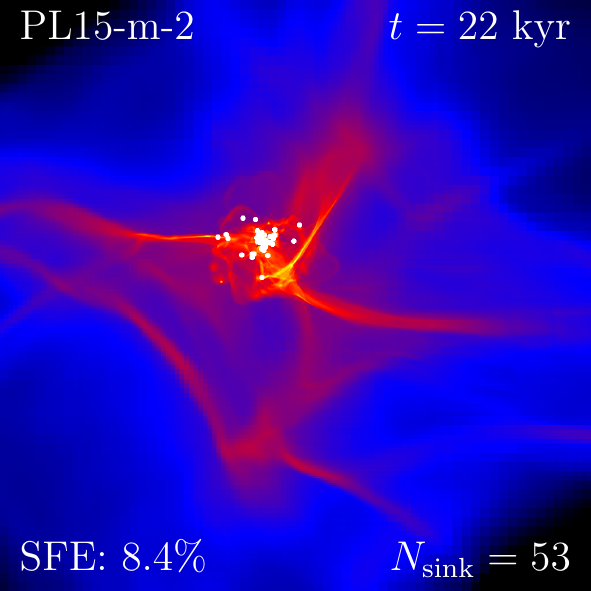}\\
    \includegraphics[width=5.0cm]{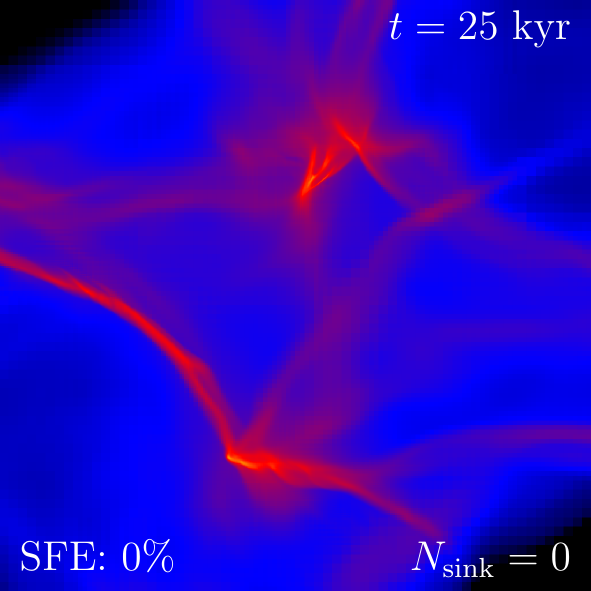}
    \includegraphics[width=5.0cm]{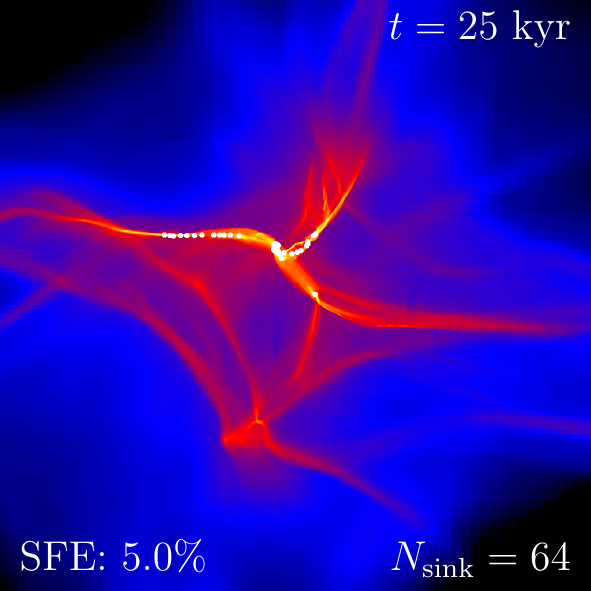}
    \includegraphics[width=5.0cm]{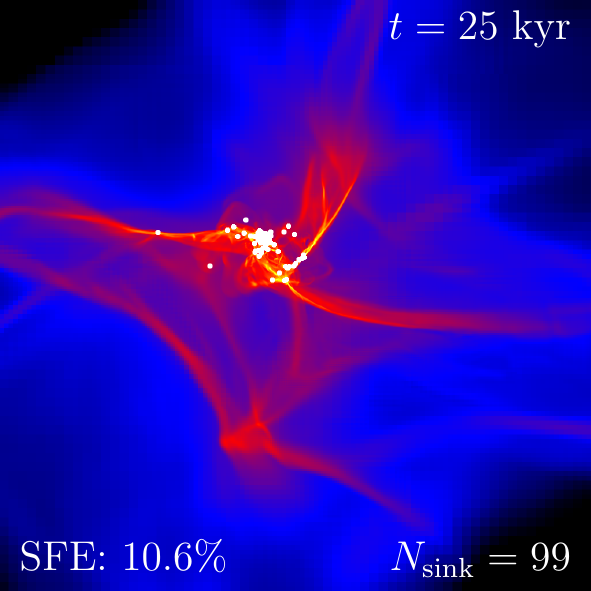}\\
    \includegraphics[width=5.0cm]{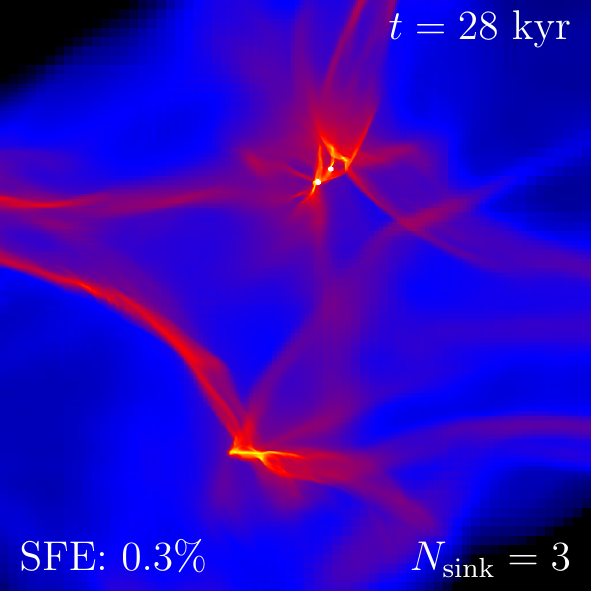}
    \includegraphics[width=5.0cm]{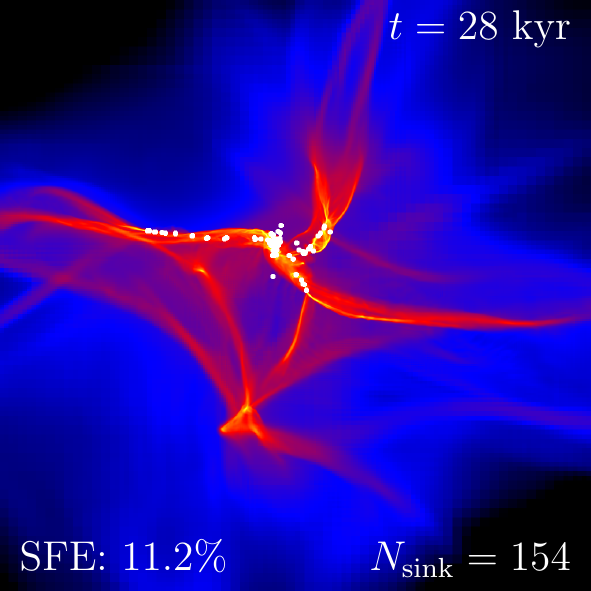}
    \includegraphics[width=5.0cm]{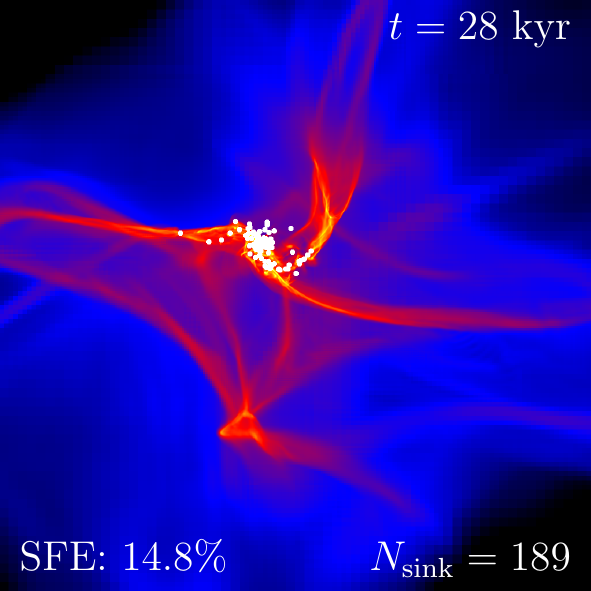}\\
    \includegraphics[width=5.0cm]{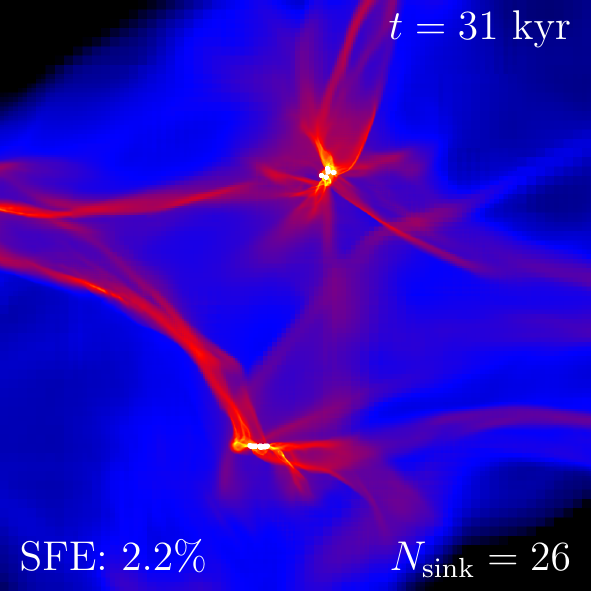}
    \includegraphics[width=5.0cm]{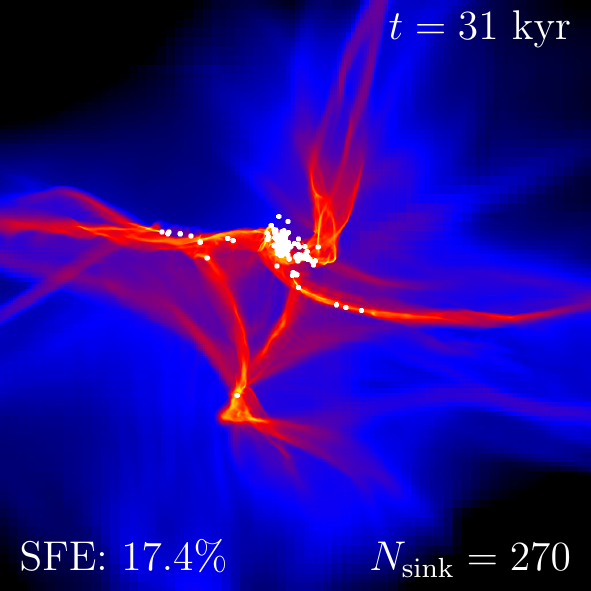}
    \includegraphics[width=5.0cm]{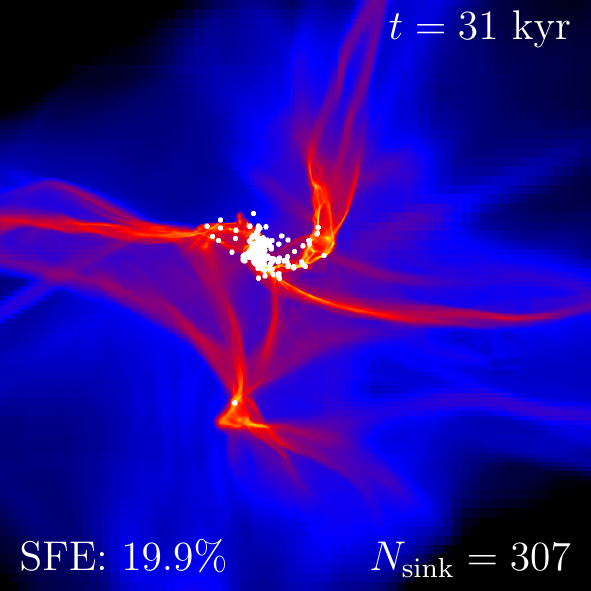}\\
    \ColorBar
  \end{minipage}
  \caption{Column density plots for the TH, BE, and PL15 profile with the velocity field m-2 in a box of $0.13$~pc in $x$ and $y$ direction. The TH-m-2 clearly develops two subclusters by the end of the simulation. The BE and PL15 runs show a similar general cloud structure that is dominated by central collapse. In the BE case the flatter initial density forms sink particles far away from the centre, whereas in the PL15 run the cluster is more compact.}
  \label{fig:column-density-plots-m-2}
\end{figure*}

A time evolution for turbulent field m-2 and the density profiles TH, BE, and PL15 is shown in figure~\ref{fig:column-density-plots-m-2}. Each row shows the column density at the same simulation time. The columns correspond to the different density profiles. The much slower central collapse in the TH case allows the formation of two distinct over-dense regions, shown at $t=22~\text{kyr}$. At that time the BE profile has formed a few stars along the long main filament. The PL15 profile has already formed more than 50 sink particles very close to each other that interact very strongly and disturb the central filamentary structure. $3$~kyr later the BE sphere formed more stars mainly along the outer arms of the main filament. Although the number of sink particles is larger than in the PL15 case at the previous time snapshot and the total mass captured in sink particles is roughly comparable, the cluster is not dominated by the gravitational attraction and $N$-body dynamics of the stars. The initial gas structure remains unperturbed. Another $3$~kyr later the TH profile eventually developed collapsing regions in completely disconnected areas. By that time the BE cluster begins to show dynamical interactions. In the last time snapshot the overall cloud structure as well as SFE and the number of sink particles is comparable for the BE and the PL15 case.

Concerning the formation of sink particles, a clear distinction between the power-law profiles and the profiles with a flat core has to be made. The power-law profiles with their high density core form a sink particle very early due to the fast collapse of the central region. In the PL20 profile and in two of the PL15 profiles, this particle remains the only particle formed in the entire simulation time. PL15 runs with more than one sink particle form them with a large time gap after filamentary structures have formed and collapse. In the BE and TH profiles this central particle does not exist, and all particles form in filaments. This different behaviour can be seen in the mass evolution (figure~\ref{fig:comparison-density-profiles}). The runs with PL15 profile form a sink in the centre right after the start. The mass therefore evolves similarly at the beginning. For the BE sphere and the uniform density distribution, the different realisations of the turbulence lead to different filamentary structures and thus influence the point in time when sink particles are created. Therefore, the mass evolution of the different simulations show large offsets (figure~\ref{fig:comparison-density-profiles}).

In general, all setups result in high total accretion rates onto the sink particles of $\dot{M}\sim1-2\times10^{-3}~M_\odot~\text{yr}^{-1}$. Only PL15-m-2 and PL15-s-2 (see detailed discussion below) show somewhat smaller values of the accretion rate. The fluctuations around the mean value strongly depend on the number of particles, their positions, and the resulting particle-particle interactions as well as accretion shielding effects. The PL20 as well as two PL15 runs only form one sink particle and show a very smooth accretion rate with small fluctuations. The accretion rates in the TH and BE profiles are influenced by the particle movements but as the clusters are not that compact the interactions are less intense. The overall similarity of the accretion rates can also be seen in the similar slope of the mass function in the upper panel of figure~\ref{fig:comparison-density-profiles}.

\subsection{Analysis of the TH Profile}
The uniform density distribution has much less mass within the core region compared to the concentrated profiles (see table~\ref{tab:instability}), and its core free-fall time is longer. The initial supersonic velocity field has time to develop significant over-densities before the global collapse becomes dominant. Therefore the evolution of the cloud at the beginning of the simulation is dominated by the turbulent motion rather than the central collapse. The turbulence crossing time and the free-fall time of the core are similar ($t_\text{tc}^\text{core}/t_\text{ff}^\text{core}$=1.64) which leads to the formation of over-dense regions all over the simulation box. These over-dense regions are very massive and evolve to locally collapsing filaments in which the first sink particles form. Filaments that are close to each other merge into sub-cores in which subclusters build up, before the central collapse sets in. After roughly one free-fall time, 20\% of the mass is collapsed into sink particles.

\begin{figure}
  \centering
  \includegraphics[width=8cm]{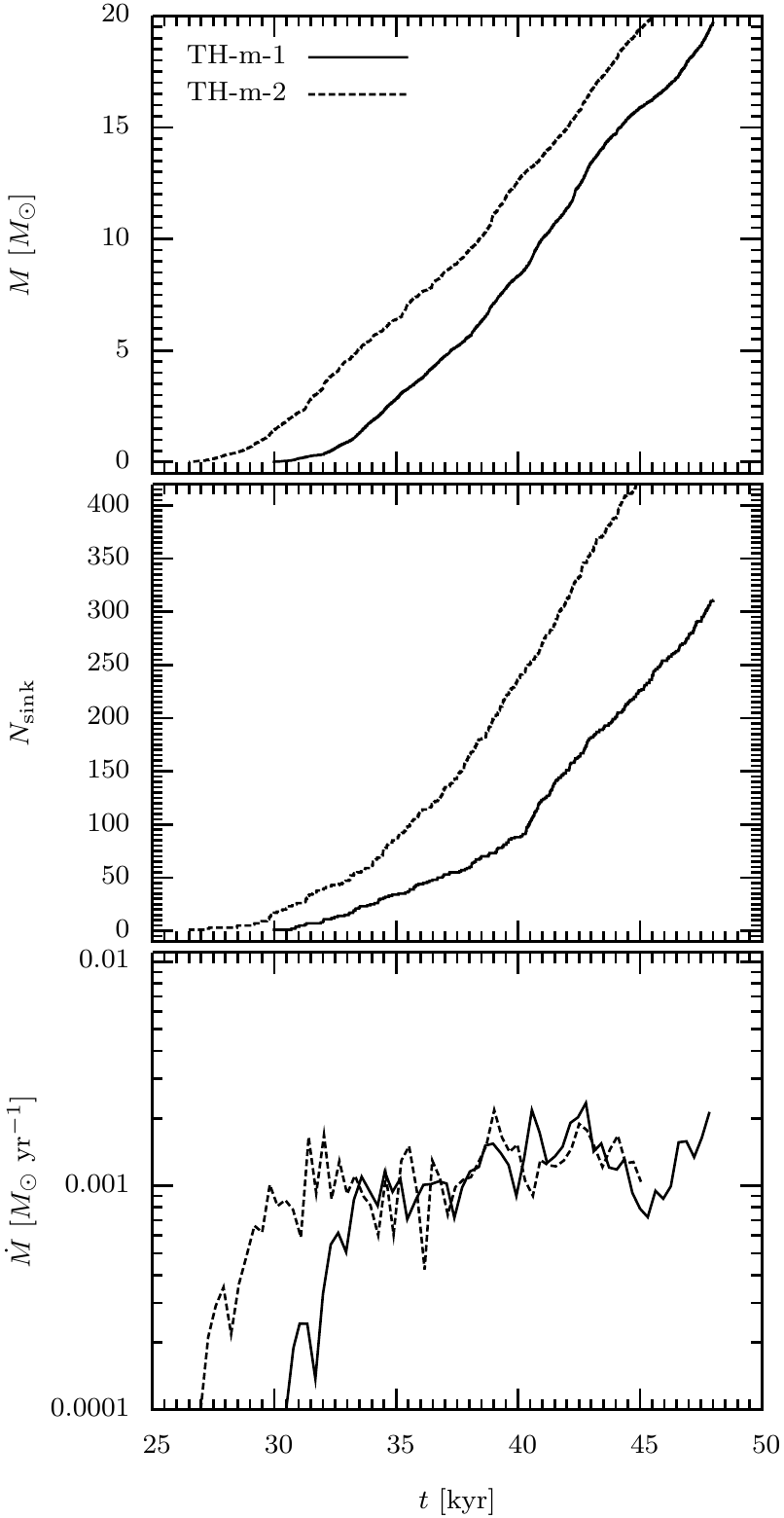}
  \caption{Sink particle evolution for the TH runs. Once sink particles have started to form the total accretion rate (lower plot) quickly reaches a value around $\dot{M}\sim10^{-3}~M_\odot~\text{yr}^{-1}$, fluctuating by a factor of roughly 2. Therefore the evolution of the mass captured in sink particles as a function of time looks very similar for both runs (upper plot), just shifted by $3-4$~kyr.}
  \label{fig:TH-particle-evolution}
\end{figure}

The accretion rate for every single sink particle is a strongly varying function with time. However, the global rate for the sum of all sink particles quickly reaches a saturated value of $\dot{M}\sim10^{-3}~M_\odot~\text{yr}^{-1}$ (figure~\ref{fig:TH-particle-evolution}), which can also be seen in the comparable slope of the total sink particle mass as a function of time. The number of sink particles is noticeably higher for TH-m-2.

\begin{figure}
  \centering
  \includegraphics[width=8cm]{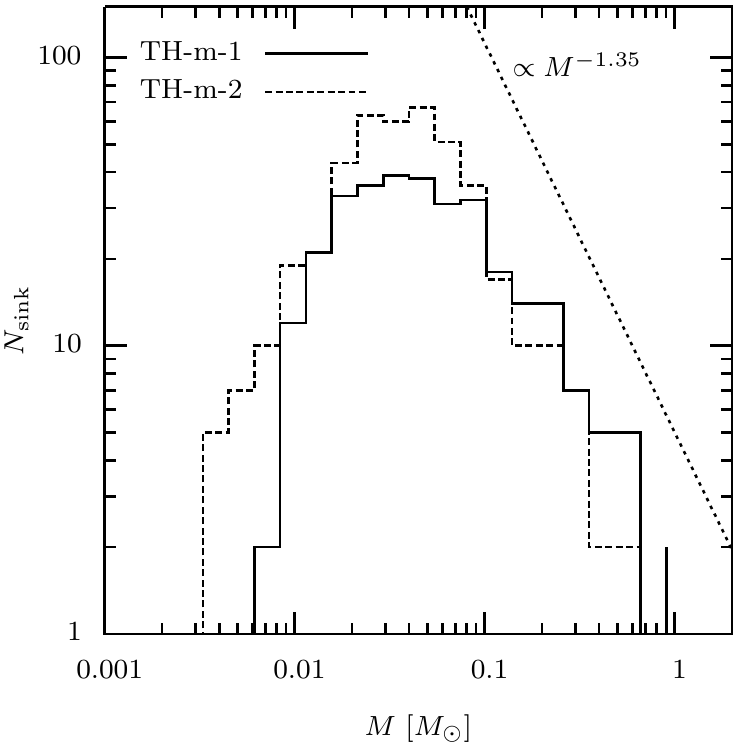}
  \caption{The mass distribution of the sink particles for the TH setup has an overall shape similar to the uniform IMF \citep{Kroupa01,Chabrier03}, but shifted to lower masses.}
  \label{fig:TH-IMF}
\end{figure}

The mass distribution of the sink particles follows an overall shape similar to the universal IMF \citep[e.g.][]{Kroupa01,Chabrier03}, but shifted to lower masses by a factor of about 10 (see figure~\ref{fig:TH-IMF}). A comparison with analytic models of the IMF \citep[e.g.,][]{Padoan02,HennebelleChabrier2008} is planned in a future contribution. Here the main conclusion is that the formation of massive stars is very unlikely in a cloud with $100~M_\odot$ and a uniform density distribution.

Since refinement is initiated in a very space-filling fashion for the uniform density distribution of the TH runs and thus computational cost became prohibitive, we only ran mixed turbulence runs with two different seeds. It should be noted, however, that the influence of the different mixtures (compressive versus solenoidal) of the turbulence has the biggest influence on the evolution and structure of the forming clusters and subclusters in the TH profiles, because TH profiles provide the most time for the turbulence to influence the cloud structure before the global collapse sets in. This will be addressed in a separate paper.

\subsection{Analysis of the BE Profile}
\begin{figure*}
  \begin{minipage}{160mm}
    \centering
    \includegraphics[width=5.0cm]{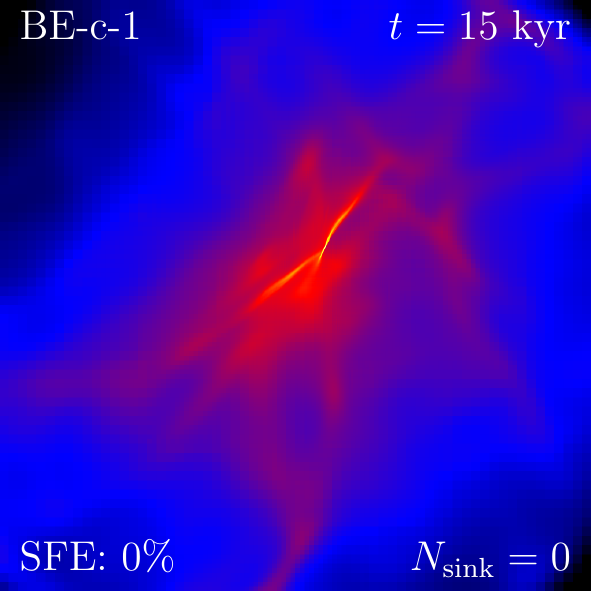}
    \includegraphics[width=5.0cm]{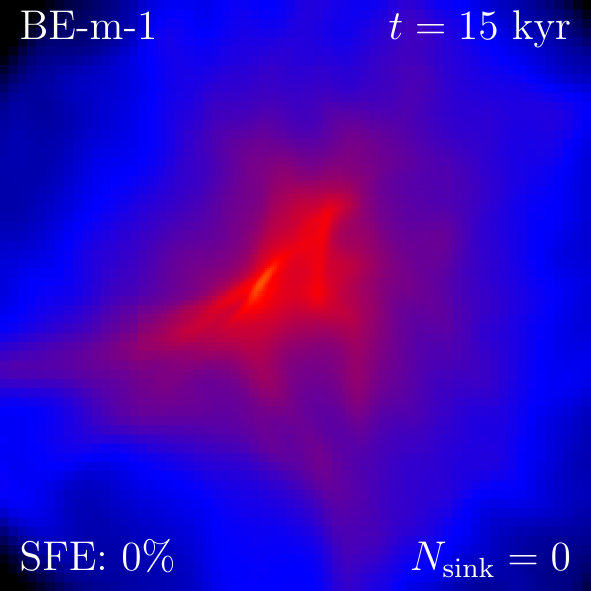}
    \includegraphics[width=5.0cm]{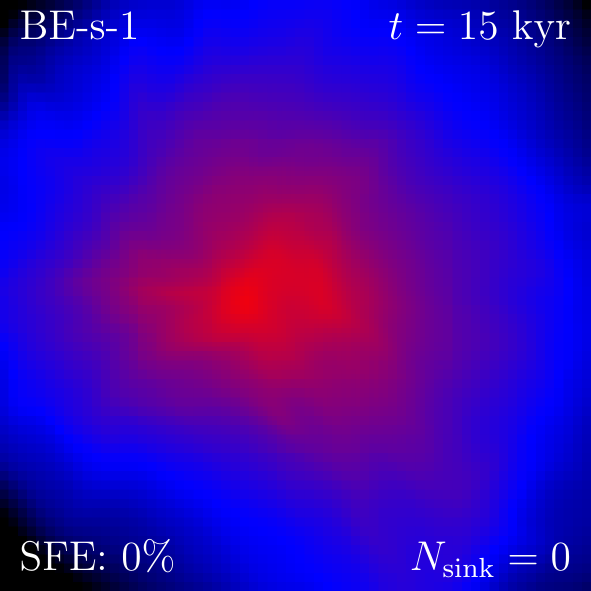}\\
    \includegraphics[width=5.0cm]{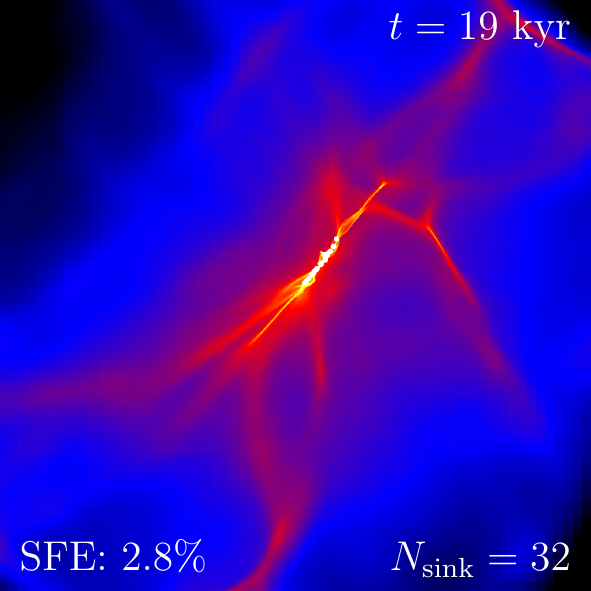}
    \includegraphics[width=5.0cm]{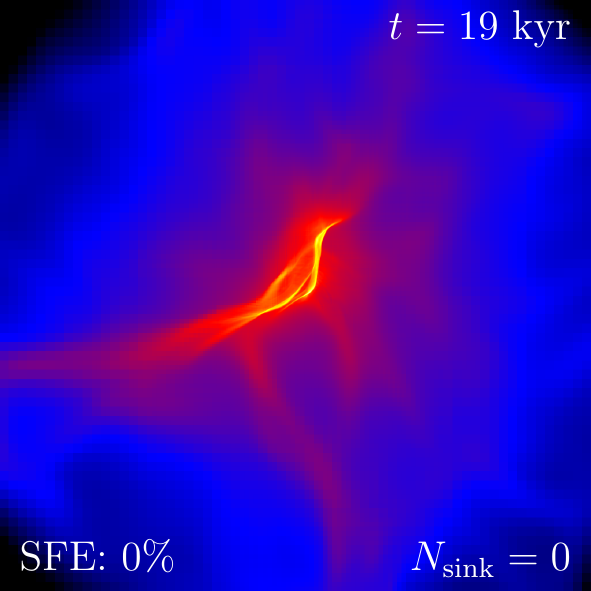}
    \includegraphics[width=5.0cm]{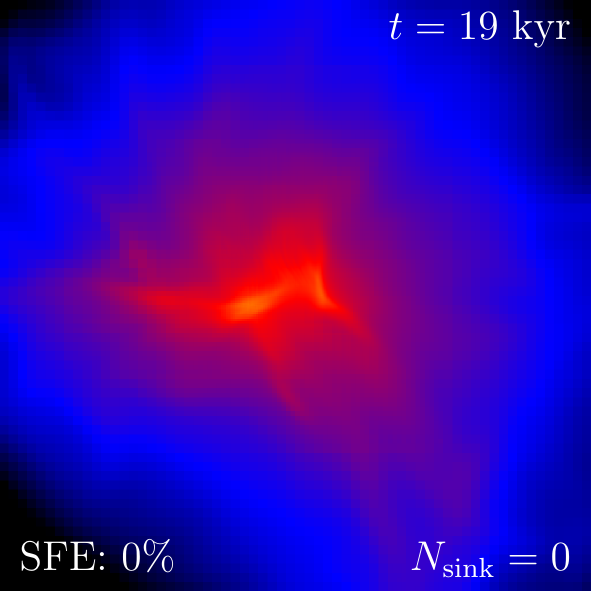}\\
    \includegraphics[width=5.0cm]{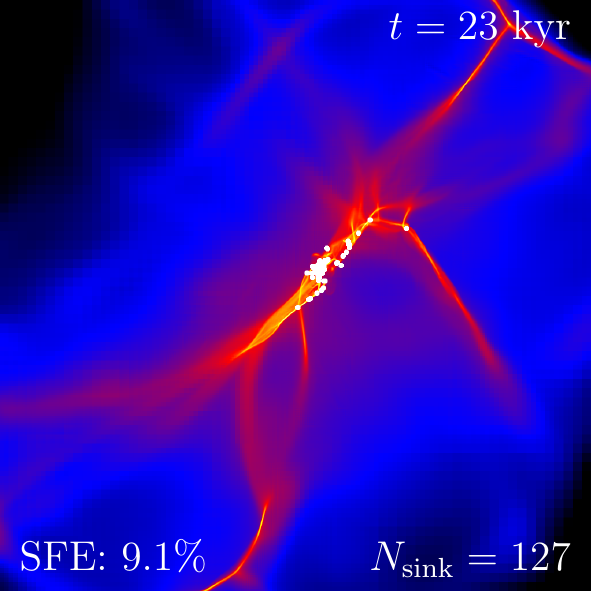}
    \includegraphics[width=5.0cm]{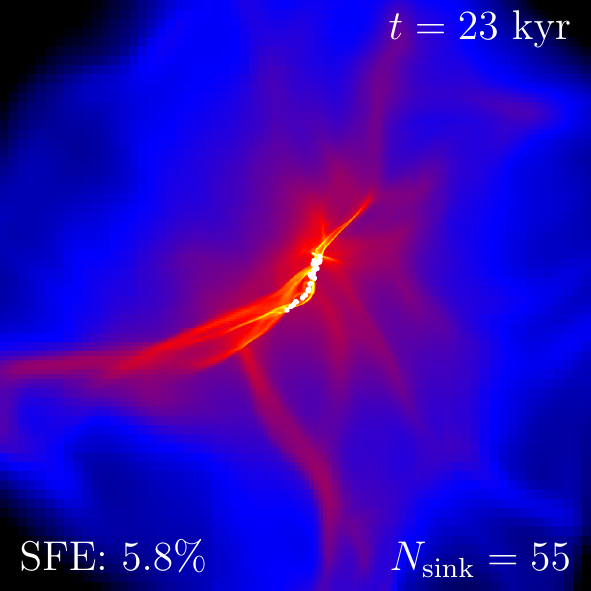}
    \includegraphics[width=5.0cm]{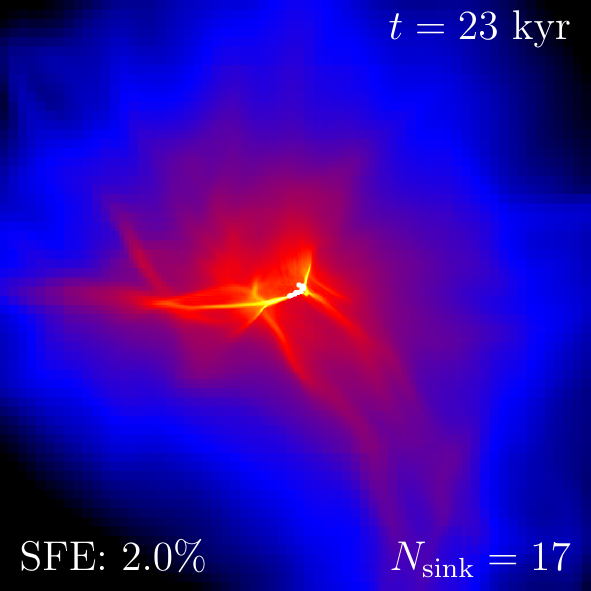}\\
    \includegraphics[width=5.0cm]{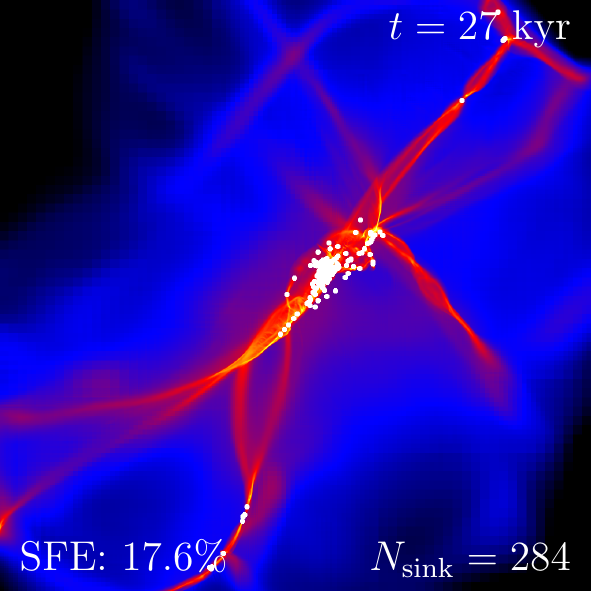}
    \includegraphics[width=5.0cm]{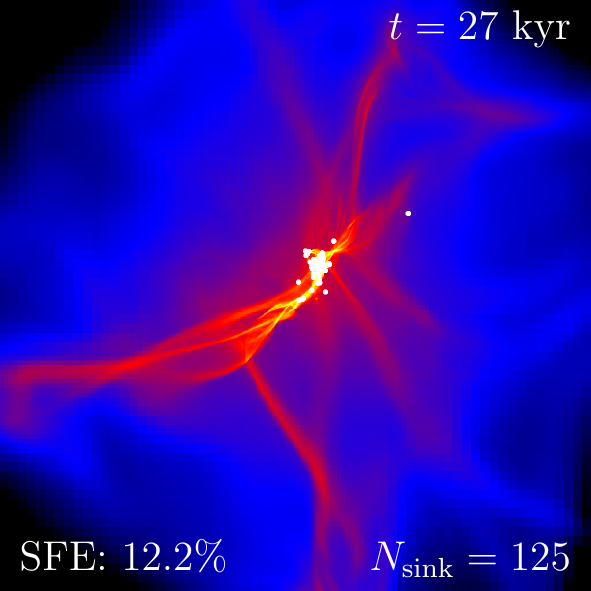}
    \includegraphics[width=5.0cm]{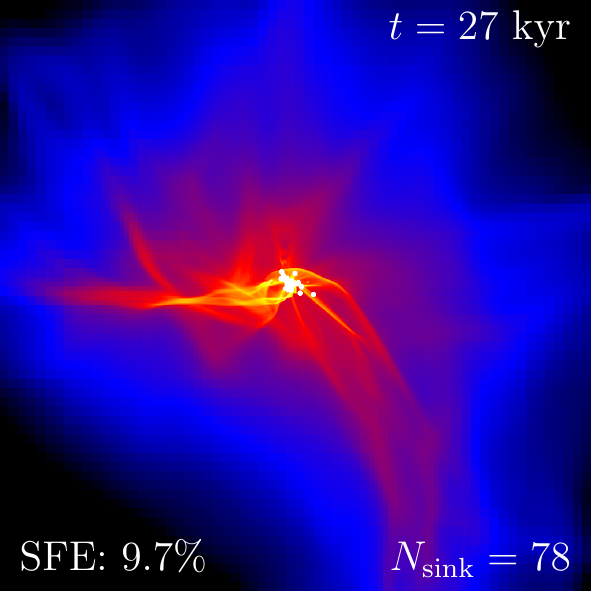}\\
    \ColorBar
  \end{minipage}
  \caption{BE column density plots for the BE density profile and three different turbulent fields. The columns show snapshots with c-1, m-1, and s-1 velocities (from left to right) for the same physical time. The box shown spans $0.13$~pc in $x$ and $y$ direction.}
  \label{fig:BE-column-density-plots}
\end{figure*}

Here the cloud evolution at the beginning is similar to the collapse of the TH core. The turbulence can form strong filaments spread over large regions of the domain. However, the different radial mass distribution leads to low-mass filaments in the outer regions. This results in a stronger central acceleration, which causes the filaments to merge near the centre of mass. The formation of large subclusters is suppressed compared to the case of the uniform density distribution. By far, most of the sink particles, which are roughly as numerous as in the TH simulations, are formed in the core region. The time evolution of the cloud for different turbulent modes with the same random velocities can be seen in figure~\ref{fig:BE-column-density-plots}. The compressive modes lead to sink particle formation about 25\% earlier than the mixed and solenoidal modes.

\begin{figure}
  \centering
  \includegraphics[width=8cm]{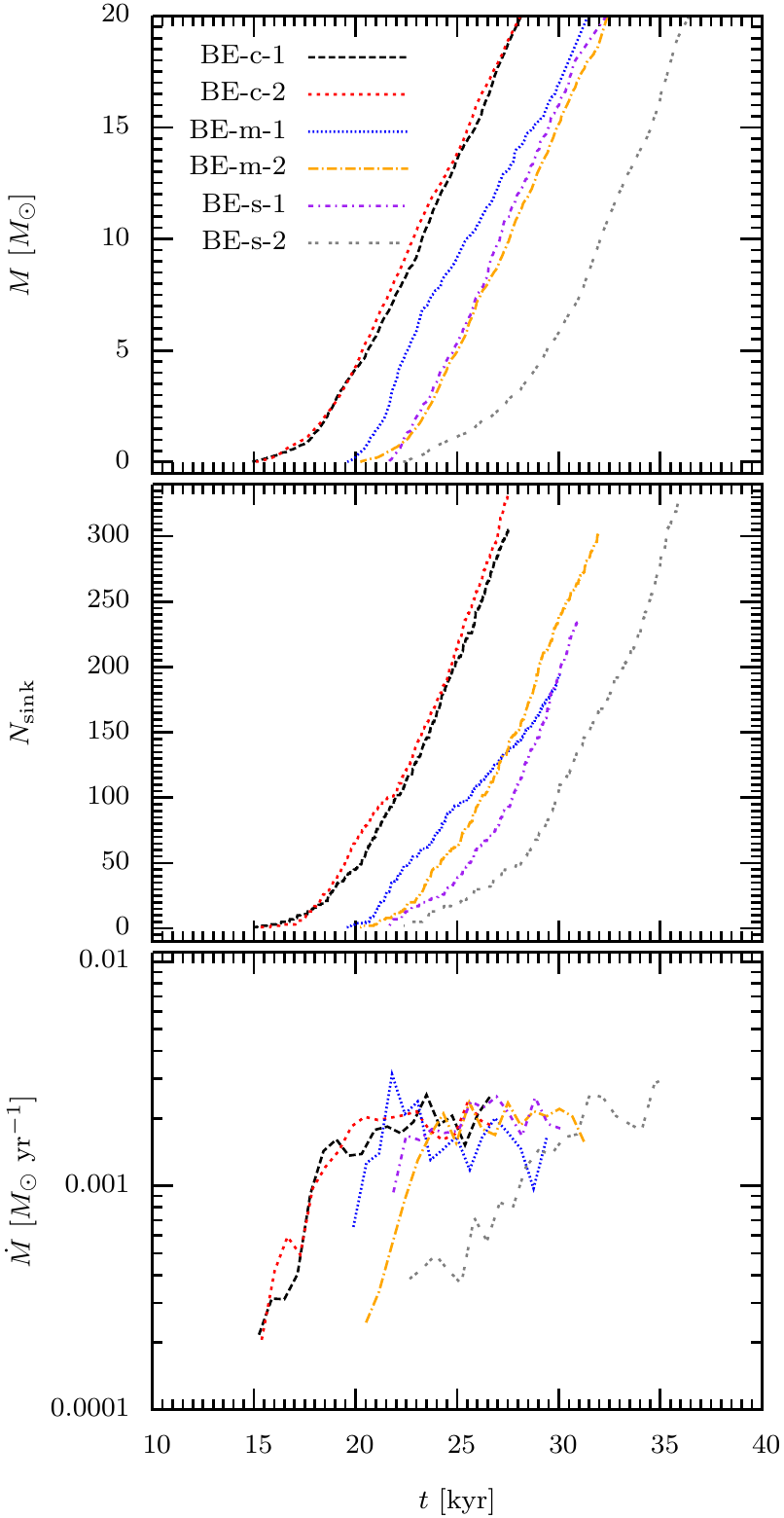} 
  \caption{Sink particle evolution in the BE runs. The upper plot shows the total mass captured in sink particles. The compressive fields form sink particles first, the mixed and solenoidal velocity fields a few kyr later. After the formation of the first sink particle the accretion rate (lower plot) approaches a value of $\sim2\times10^{-3}~M_\odot~\text{y}^{-1}$, independent of the initial turbulent field. The number of sink particles also shows a similar evolution for all setups (central plot).}
  \label{fig:BE-particle-evolution}
\end{figure}

The time evolution of the global sink particle properties are shown in figure~\ref{fig:BE-particle-evolution}. Although the random seed strongly determines the location and orientation of the filaments, the particle formation between BE-c-1 and BE-c-2 is almost indistinguishable. In the case of mixed and solenoidal modes the choice of the random seed significantly changes the time at which sink particles form. However, after the creation of sink particles has set in, the particle production rate with time as well as the total mass accretion rate is quite similar for all runs, not reflecting the structure of the initial turbulence at all. Only the BE-s-2 setup needs some more time until it reaches the asymptotic value of $\dot{M}\sim2\times10^{-3}~M_\odot~\text{y}^{-1}$.
However, the accretion rate of individual sink particles varies strongly with time.
\begin{figure}
  \centering
  \includegraphics[width=8cm]{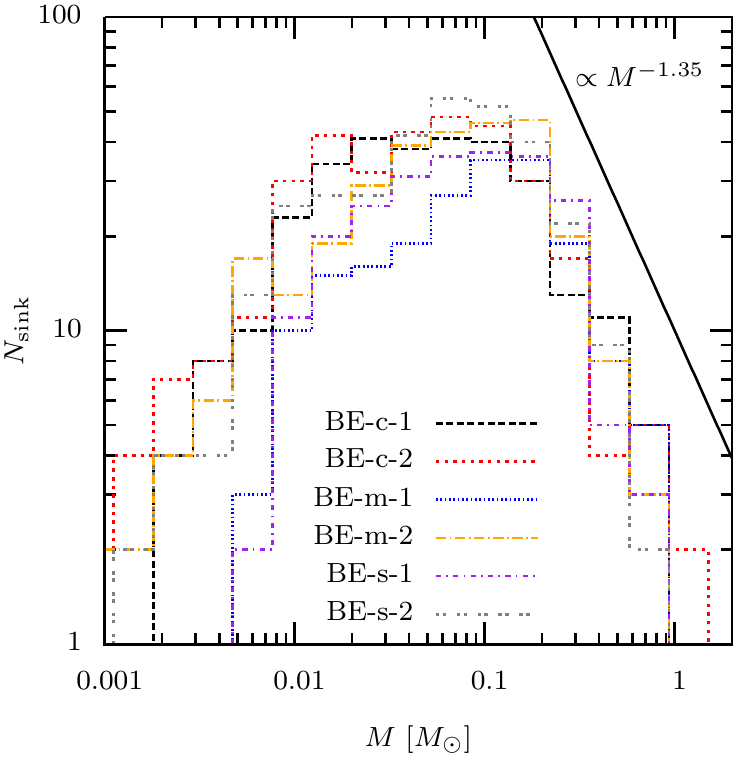}
  \caption{IMF for the BE setups. For all turbulent setups the IMF looks very similar. The distribution function mainly follows the general shape of a uniform IMF, but with lower average masses.}
  \label{fig:BE-IMF}
\end{figure}

The mass distribution of the sink particles (figure~\ref{fig:BE-IMF}) also shows a typical IMF structure like the TH runs, also shifted to much lower masses. This leads to the conclusion that the stronger central density concentration and the resulting stronger in-fall properties are still way too inefficient in forming massive stars. 

\subsection{Analysis of the PL15 Profile}

\begin{figure}
  \centering  
  \includegraphics[width=8cm]{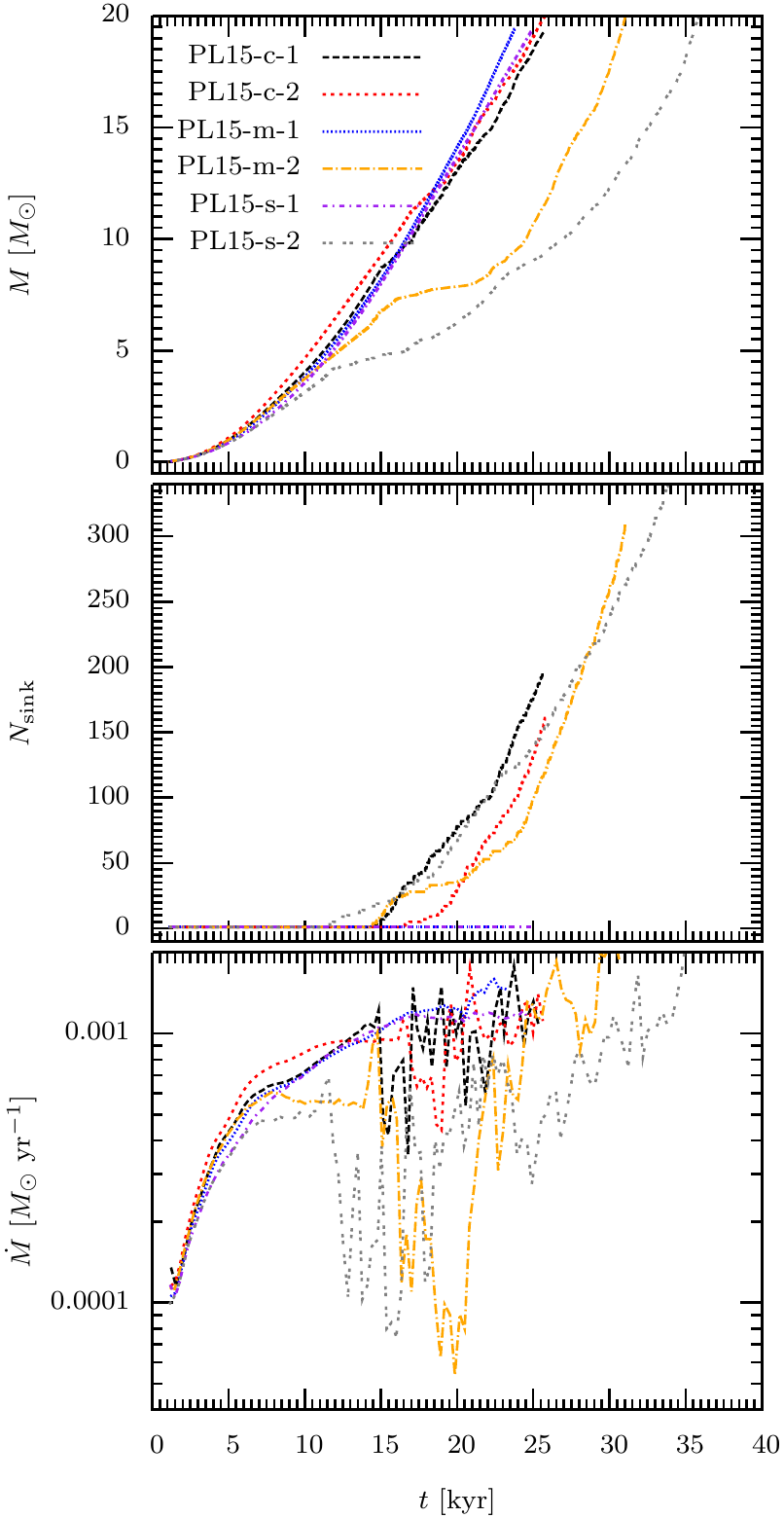}
  \caption{PL15 particle evolution. The upper plot shows the total mass captured in sink particles. Apart from the m-2 and s-2 velocity field, the mass evolution is very similar. This can also be seen in the lower plot, showing the accretion rate. In case of more sink particles, the accretion rate varies much more strongly with time. This is due to strong particle-particle interactions in the compact stellar cluster. If the cloud fragments and collapses in different regions the number of stars follows similar curves (central plot). However, the total number of particles differs much more than in other density setups.}
  \label{fig:PL15-particle-evolution}
\end{figure}

From the very beginning of the simulation, the PL15 profiles show a considerably different evolution compared to the TH and BE profiles. Due to the strong mass concentration, the first sink particle forms close to the centre very early, after roughly $1~\text{kyr} \approx 0.06~t_\text{ff}^\text{core}$. The formation of this sink particle is not influenced by extended filaments, because the formation time of filaments is much larger than the time for central collapse. The central particle has a high and smooth accretion rate in all PL15 runs, which allows it to grow to the most massive sink particle in the simulation, while filaments in the outer regions start to form later (figure~\ref{fig:PL15-particle-evolution}). Whether secondary sink particles form strongly depends on the random seed of the turbulence, as well as on the nature of the modes. All simulations with compressive modes lead to the formation of many sink particles in the filaments. On the other hand, mixed and solenoidal modes lead to either one (PL15-m-1 \& PL15-s-1) or a few hundred particles (PL15-m-2 \& PL15-s-2). A possible explanation for this dichotomy could be the influence of tidal forces, which can suppress the growth of the initial perturbations induced by the turbulence. In a density profile steeper than $r^{-1}$ (see appendix~\ref{sec:tidal_forces}), tidal forces start to shear radial density fluctuations apart, thus reducing the chance of initial perturbations to grow by self-gravity. For the BE profile the central region of the cloud has a shallower density profile than $r^{-1}$, the PL15 profile a slightly steeper one. Over-dense regions that can marginally grow in the BE profile may be sheared apart in the corresponding PL15 profile with the same velocity field. However, the turbulence is supersonic and the density power-law exponent is not far from the critical one. This is why different locations and strengths of converging and diverging regions of the velocity field may easily overcome the shearing effect and cause the big differences between PL15-m-1/PL15-s-1 and PL15-m-2/PL15-s-2. Indeed, an analysis of the density-weighted divergence of the initial velocity fields shows that seed 2 produces stronger compressions in regions of high density than seed 1. Taken together with the fact that fragmentation into multiple objects always occurs for the purely compressive fields, this shows the importance of compressive modes for triggering the formation of dense fragments.

In the first $10~\text{kyr}$, the evolution of all PL15 simulations is quite similar. During that time all simulations have only formed one central sink particle. As soon as other sink particles form, the situation changes significantly. In the case of multiple sink particles, their particle-particle interactions in the stellar cluster disturb the central in-fall and redirect the central gas velocities.

\begin{figure}
  \centering
  \includegraphics[width=8cm]{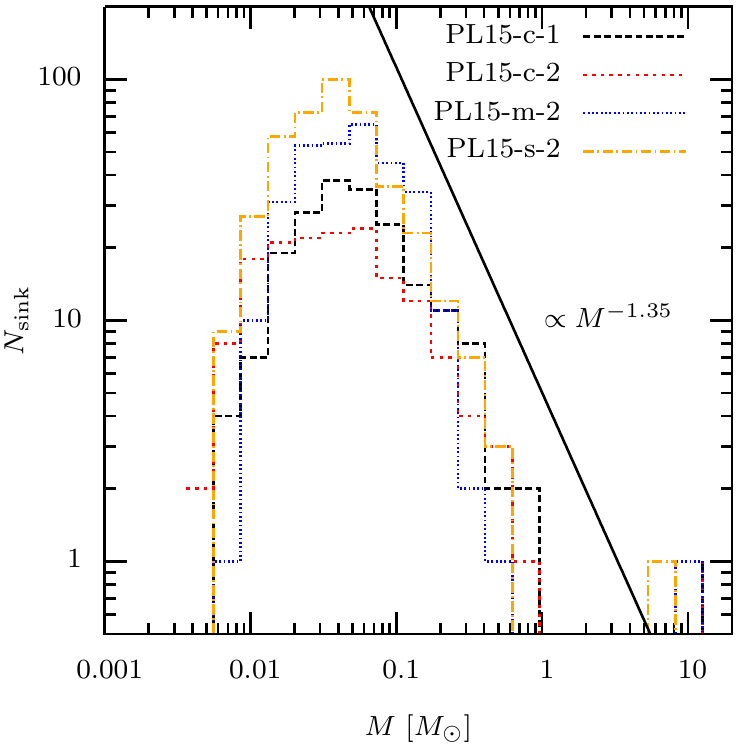}
  \caption{IMF for the PL15 runs. All runs form one very massive sink particle, which is by far the most massive one in the cluster, indicated by the single peak around $10~M_\odot$. The continuous set of low-mass particles below the mass gap shows similarities with the universal IMF, again shifted to almost 10 times lower masses as in the TH and BE profiles.}
  \label{fig:PL15-IMF}
\end{figure}

Although the total number of sink particles as a function of time is similar for PL15-c-1, PL15-c-2, PL15-m-2, and PL15-s-2, their spatial distribution differs between the runs with compressive velocity field (PL15-c-1, PL15-c-2) and the runs PL15-m-2 and PL15-s-2 with mixed and solenoidal fields. In the former, the sink particles are located in filaments much farther away from the centre, resulting in weaker particle-particle interactions and allowing the particles to remain located in their dense parental filament. The runs PL15-m-2 and PL15-s-2 are dominated by the in-fall of less centrally located and hence less massive filaments. The local gravitational collapse inside these filaments is therefore delayed until the filament approaches the dense core. Sink particles show much lower mean separations which increases the strength and impact of particle-particle interactions. The induced cluster dynamics reduces the total mass accretion rate because individual sinks stop accreting if they are kicked out of the dense gas regions. This effect can also be seen in the IMF (figure~\ref{fig:PL15-IMF}). PL15-m-2 and PL15-s-2 have many more sink particles, but the final mass of the central one is lower than in the runs with compressive fields (see table~\ref{tab:Nsinks_and_simtime}). Hence, the accretion onto the central object is starved by the fragmentation around it \citep{Peters10a}

\subsection{Analysis of the PL20 Profile}
\begin{figure}
  \centering
  \includegraphics[height=7.8cm]{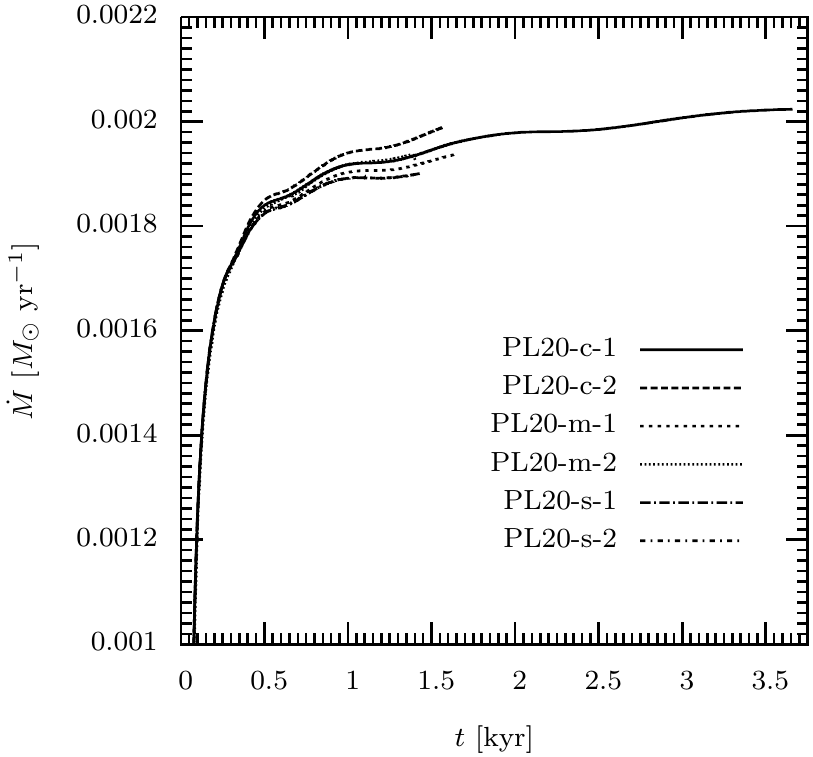}
  \caption{Comparison of sink particle accretion rate for the PL20 profile with different turbulent velocity fields. Note that the accretion rate is plotted in linear scale in order to see the small differences between the runs with different turbulence. In all cases only one sink particle is created over the whole simulation time.}
  \label{fig:PL20-turb-comp}
\end{figure}

For the PL20 density profile with the compressive turbulent field, only one sink particle was created already after $0.13~\text{kyr}$ which is only $0.012~t_\text{ff}^\text{core}$. As this velocity field is the most likely one to form more than one sink particle, the other turbulence realisations are not simulated entirely. This density profile is gravitationally too unstable for the turbulence to have an impact on the density evolution and the fragmentation of the gas sphere within a core free-fall time. As the turbulence crossing time is about 20 times longer than the core free-fall time, the small influence of the turbulence is expected. The accretion rates for all realisations of this setup are very similar (figure~\ref{fig:PL20-turb-comp}). Therefore only the setup with compressive mode~1 (PL20-c-1) was simulated up to a star formation efficiency of 20\%. In conclusion, a $\rho(r)\propto r^{-2}$ density profile does not reproduce a realistic IMF but helps to form massive stars.

In order to investigate the threshold turbulent energy that is needed to cause other regions to fragment and collapse besides the central region, three additional PL20 profiles with higher velocities were investigated (see tab.~\ref{tab:runs}). The turbulence in PL20-c-1b, with twice as high velocities than our standard PL20 run, is still not strong enough to significantly alter the result. There is still only one sink particle created, accreting mass at a very high rate. For PL20-c-1c with velocities four times as high as in PL20-c-1 ($\mathcal{M} = 13.1$), the situation changes. The stronger turbulence leads to the formation of other sink particles apart from the central one. However, the central particle in this run still contains 90\% of the mass ($M=18~M_\odot$) at the end of the simulation, and the second most massive particle is more than one order of magnitude less massive. Similar results are obtained from PL20-c-1d with a Mach number of $\mathcal{M} = 19.7$. More sink particles form, but still the central star is the most massive one with $M=13~M_\odot$.

\section{Discussion}
\label{sec:discussion}
Our results clearly show that diverse initial conditions lead to completely different cloud structures and collapse scenarios. However, the strong dependence of the simulation outcome on the initial conditions may be moderated by different input physics like radiation or magnetic fields and the effects due to rotation. 

Our simulations indicate that massive stars can form without the aid of radiation and magnetic fields just from choosing centrally concentrated density profiles. In contrast, our isothermal cloud setups with flat density distributions fail to produce massive stars. We note that this result could change significantly if more massive clouds with more Jeans masses are used. We find in our simulations of isothermal gas that clouds with an initially uniform density distribution tend to overproduce low-mass proto-stars and have difficulty forming sufficient numbers of high-mass objects. This is in qualitative agreement with the simulations done by \citet{Bate03} and \citet{Bate05}. They used a uniform density distribution and solenoidal velocity fields, which seems to represent the conditions that inevitably lead to a large number of low-mass objects. This is also consistent with the calculations by \citet{Offner08}, \citet{Klessen98}, \citet{Klessen00}, \citet{KlessenHeitsch00}, \citet{Klessen01}, and \citet{HeitschMacLow01}, who tested the influence of driven and decaying turbulence in a uniform density box. In order to suppress fragmentation and/or enhance the formation of massive stars in flat density profiles, more physics may help, which is addressed in three different approaches, namely radiation feedback, magnetic fields and stellar collisions. Concerning the first process, \citet{Kratter06} derived an analytical model to address the fragmentation process in massive discs. Indeed, \citet{Bate09a}, \citet{Krumholz09}, \citet{Peters10a, Peters10b, Peters10c} found reduced fragmentation in simulations. However, radiative feedback does not suppress fragmentation entirely. Alternatively, magnetic fields tend to reduce fragmentation. \citet{HennebelleTeyssier2008}, \citet{Ziegler05}, and \citet{Buerzle10} investigated the influence of magnetic fields in low-mass cores, \citet{Banerjee09}, \citet{Peters10d}, and \citet{Hennebelle11} noted reduced fragmentation in high mass cores. But again, the fragmentation is not fully suppressed. The formation of massive stars by stellar collisions was proposed by \citet{Zinnecker07}. However, \citet{Baumgardt10} showed that under realistic cloud conditions the contribution of stellar collisions can be neglected.

In earlier studies, the discussion about cloud fragmentation and the formation of massive stars is strongly focused on the physical processes, not taking heed of the importance of initial conditions. As the time scale for star formation is of the order of a few dynamical times \citep{Ballesteros99b, Elmegreen00, Hartmann01, MacLow04, Elmegreen07}, the star-forming core has only little time to interact with the surrounding medium. The boundary and initial conditions are therefore decisive key properties for the collapse scenario and the star formation outcome. To fully understand the formation of a star cluster therefore requires knowledge of both the initial conditions for the cluster-forming cloud core (density profile, temperature, turbulent velocity content) as well as the time-dependent boundary conditions (as the core is connected to the overall turbulent cloud environment and may grow in mass by accumulation of gas at the stagnation points of larger-scale convergent flows).

Given the sensitivity of the dynamical evolution on the choice of the initial density profile, it is of pivotal importance to seek guidance from observations. On small scales ($\ll1~\text{pc}$) the observed cores clearly deviate from a uniform density \citep[e.g.,][]{Pirogov09, Koenyves10, Bontemps10}. The outer regions of molecular cloud cores can be described by a power-law with $\rho\propto r^{-1.5}$. In the centre of a dense core, however, the approach of a power-law function seems to be inconsistent with observations, which identify the central region of the core to be flat \citep{Motte98, Ward-Thompson99}. Starless cores may often be fitted with a critical Bonnor-Ebert sphere, cores with stars are often better fitted with super-critical ones \citep{Teixeira05, Kandori05, Kirk05}. \citet{Krumholz07, Krumholz10} use very similar setups to our PL15 density profile, emphasising the importance of radiative feedback for the formation of massive stars. In this density profile the central region inevitably determines the collapse time scale and the formation of the first proto-stellar object. Our current analysis indicates that following a power-law profile to very small radii ($< 10^3$~AU) introduces a bias towards forming a massive central object without much fragmentation around it. Adding radiative feedback does not change the outcome significantly in view of the very short central collapse time scales.

We can also look at the way ISM turbulence is treated in other numerical studies. \citet{Bate03}, \citet{Bate05}, \citet{Bate09, Bate09a, Bate09b}, \citet{Bonnell03, Bonnell04}, and \citet{Bonnell05}, for example, always used divergence-free, decaying turbulent fields. \citet*{Clark08}, \citet*{Clark08a}, \citet{Offner08}, and \citet{Krumholz07} do not specify the nature of the modes they select for their turbulence. As our results show that compressive, decaying modes lead to significantly earlier collapse and more elongated, shocked structures in the flat density profiles (TH and BE) than purely solenoidal turbulence, this is an important aspect of the star formation process that deserves further consideration. A systematic study of different modes of the turbulence was done by \citet{Federrath08,Federrath09,Federrath10b} and \citet{SeifriedEtAl2010}, but in a periodic box with driven turbulence and without gravity. These studies find the expected trend that compressive modes initiate faster collapse and higher accretion rates than purely solenoidal turbulence. However, the influence of the different modes is stronger in driven turbulence with self-gravity than in the decaying turbulence runs analysed here. Since dense cores are typically embedded in large-scale, turbulent molecular clouds, an effective driving of the internal turbulence from outside the core is expected \citep[e.g.,][]{Klessen00b, Federrath10b}.

\section{Summary and Conclusions}
\label{sec:summary_conclusion}
We performed a parameter study of the fragmentation properties of collapsing isothermal gas cores with different initial conditions. We combined four different density profiles (uniform, Bonnor-Ebert type, $\rho\propto r^{-1.5}$, and $\rho\propto r^{-2}$) and six different turbulent, decaying velocity fields (compressive, mixed, and solenoidal, each with two different random seeds). For these simulations we neglected radiation, magnetic fields, and initial rotation, in order to study the direct influence of the initial density profile and the character of the turbulence. The cloud evolution as well as the star formation and their properties were examined. Here we list our main conclusions:

The density profile strongly determines the number of formed stars, the onset of star formation, the stellar mass distribution (IMF), and the spatial stellar distribution.
\begin{itemize}
\item Flat profiles (uniform density and Bonnor-Ebert profiles) produce many sink particles in elongated filaments. The formation of sink particles starts after slightly more than half of a core free-fall time for the uniform cloud and after roughly one core free-fall time for the Bonnor-Ebert setups. The runs with initially uniform density produce subclusters in merging filaments in outer regions of the cloud. Even the relatively weak mass concentration in the centre of the Bonnor-Ebert setups suppresses the formation of subclusters. Both density profiles show an initial mass function with the high-mass end consistent with the Salpeter slope. In the case of initial compressive velocity fields, star formation sets in 25\% earlier than in the solenoidal case. The mixed turbulent fields are in between the two extreme cases.
\item The $\rho\propto r^{-1.5}$ profiles always form one sink particle in the centre of the cloud at an early stage. This sink particle accretes gas at rate of $\sim10^{-3}~M_\odot~\text{y}^{-1}$ and grows to the most massive particle by far. The formation of unstable filaments depends sensitively on the initial turbulent field. The formation of additional sink particles only occurs after a time delay of $\sim0.3~t_\text{ff}$. The mass distribution of these sink particles shows a high-mass slope consistent with the Salpeter slope, but has a wide gap between this mass continuum and the central massive star of almost an order of magnitude in mass. The spatial distribution shows a compact structure around the centre of the cloud and no subclustering. The column density of the filamentary structure looks extremely similar for a $\rho\propto r^{-1.5}$ run and the corresponding Bonnor-Ebert run with the same turbulent field, not reflecting the significantly different stellar properties.
\item The $\rho\propto r^{-2}$ density profile quickly leads to the formation of one single, central sink particle. The formation of other stars is strongly inhibited due to the rapid collapse compared to the time scale for filament formation. In this scenario further star formation can only be triggered by higher Mach numbers of the turbulence, if the ratio of turbulent energy to gravitational energy is increased to about unity.
\end{itemize}

The realisation of the turbulent velocity field has a major impact in the different morphology of the filamentary structure, their orientation, and shape.
\begin{itemize}
\item In the uniform density profile the random seed of the velocity determines the position of filaments from which stars form, and thus the location of the stellar subclusters. In addition, the number of sink particles generally depends on the random seed of the turbulence. Similar results are obtained for the BE profile.
\item The $\rho\propto r^{-1.5}$ profile, which marks the transition between one central massive sink particle and many low mass ones, is very sensitive to the random seed. Different realisations may switch between one single star and several hundred. The formation time and location of the central, first sink particle, however, is not influenced by the random seed.
\item The $\rho\propto r^{-2}$ setups are not noticeably influenced by the turbulence. The short collapse time of the core compared to the turbulence crossing time does not allow for turbulence to strongly influence the evolution.
\end{itemize}

Our results suggest that massive stars predominantly form out of highly unstable cloud cores which are either strongly centrally concentrated or much more massive than modelled here, allowing stars to accrete form a larger mass reservoir. The density configuration with $\rho\propto r^{-1.5}$ seems to be the most sensitive one concerning the number of collapsing fragments for different turbulent velocities.

Overall we conclude that the choice of the initial density profile is an extremely important, perhaps even the most important parameter determining the fragmentation behaviour of high-mass proto-stellar cores. Choosing an ideal simplified density profile strongly preordains the subsequent star cluster properties. This implies that the effects of different physical processes can only be reliably compared if the initial density profile is the same. In realistic star formation simulations, the formation of these cores needs to be taken into account and cores need to be formed self-consistently from larger clouds.

\section*{Acknowledgement}
We would like to thank the referee, Richard H. Durisen, for his very detailed and constructive comments. P.G.~acknowledges supercomputer grants at the J\"{u}lich supercomputing centre (NIC~3433) and at the CASPUR centre (cmp09-849). P.G.~and C.F.~are grateful for financial support from the International Max Planck Research School for Astronomy and Cosmic Physics (IMPRS-A) and the Heidelberg Graduate School of Fundamental Physics (HGSFP), funded by the Excellence Initiative of the Deutsche Forschungsgemeinschaft (DFG) under grant GSC~129/1. C.F., R.B.~and R.S.K.~acknowledge financial support from the Landesstiftung Baden-W\"{u}rrtemberg via their program International Collaboration II (grant P-LS-SPII/18) and from the German Bundesministerium f\"{u}r Bildung und Forschung via the ASTRONET project STAR FORMAT (grant 05A09VHA). C.F.~furthermore acknowledges funding from the European Research Council under the European Community's Seventh Framework Programme (FP7/2007-2013 Grant Agreement no. 247060). R.B.~acknowledges funding of Emmy Noether grant BA~3706/1-1 by the DFG. R.B.~is furthermore thankful for subsidies from the FRONTIER initiative of the University of Heidelberg. R.S.K.~acknowledges subsidies from the DFG under grants no.~KL1358/1, KL1358/4, KL1358/5, KL1358/10, and KL1358/11, as well as from a Frontier grant of Heidelberg University sponsored by the German Excellence Initiative. This work was supported in part by the U.S.~Department of Energy contract no.~DEAC-02-76SF00515. R.S.K.~also thanks the Kavli Institute for Particle Astrophysics and Cosmology at Stanford University and the Department of Astronomy and Astrophysics at the University of California at Santa Cruz for their warm hospitality during a sabbatical stay in spring 2010. The FLASH code was developed in part by the DOE-supported Alliances Center for Astrophysical Thermonuclear Flashes (ASC) at the University of Chicago.

\bibliographystyle{apj}
\bibliography{astro.bib}

\begin{appendix}

\section{Resolution Study}
\label{sec:resolution}
We test the influence of the numerical effective resolution of the code on the collapse by simulating the different cloud setups with different resolutions. The order of the numerical resolution was chosen such that the isothermal approximation for the equation of state is appropriate and the computational effort is acceptable. The different resolutions have acronyms corresponding to the maximum refinement level (RL) in the code: $l_\text{max}=7$ (RL07), $l_\text{max}=8$ (RL08), $l_\text{max}=10$ (RL10) and $l_\text{max}=12$ (RL12). Due to the different sizes of the smallest cell, the maximum gas density before creating sink particles, as well as the accretion radius vary. A comparison of the parameters can be seen in table~\ref{tab:resolution-parms}.

As the computational time for the BE and the TH profile are very large (i.e., more than an order of magnitude larger than for the PL20 profile, because of the quite space-filling refinement in the evolution of these profiles), these setups have only been compared in an early evolutionary stage. The highly concentrated PL20 cloud has been investigated in more detail: for a longer evolution time, for more different resolutions and analytically.

\begin{table*}
  \begin{minipage}{126mm}
    \caption{Main simulation parameters for different resolutions}
    \label{tab:resolution-parms}
    \begin{tabular}{lcccccc}
      refinement & eff.~res. & $\Delta x$ [AU] & $r_\text{accr}$ [AU]& $\rho_\text{max}$ [g~cm$^{-3}$] & $n_\text{max}$ [cm$^{-3}$]\\
      \hline
      RL07 & $\phantom{0}\phantom{0}512^3$   & $104.4 $                    & $313.3 $                    & $3.85\times10^{-16}$ & $1.01\times10^{8\phantom{0}}$ \\
      RL08 & $\phantom{0}1024^3$             & $\phantom{0}52.2$           & $156.7 $                    & $1.54\times10^{-15}$ & $4.03\times10^{9\phantom{0}}$ \\
      RL10 & $\phantom{0}4096^3$             & $\phantom{0}13.1$           & $\phantom{0}39.2$           & $2.46\times10^{-14}$ & $6.45\times10^{9\phantom{0}}$\\
      RL12 & $16384^3$                       & $\phantom{0}\phantom{0}3.3$ & $\phantom{0}\phantom{0}9.8$ & $3.94\times10^{-13}$ & $1.03\times10^{11}$\\
      \hline
    \end{tabular}
    
    \medskip
    Main simulation parameters for different effective resolutions. The accretion radius of the sink particles $r_\text{accr}$ is set to 3 times the minimum cell size $\Delta x$.
  \end{minipage}
\end{table*}

\subsection{BE Profile}
\begin{figure*}
  \begin{minipage}{180mm}
    \centering
    \includegraphics[height=8cm]{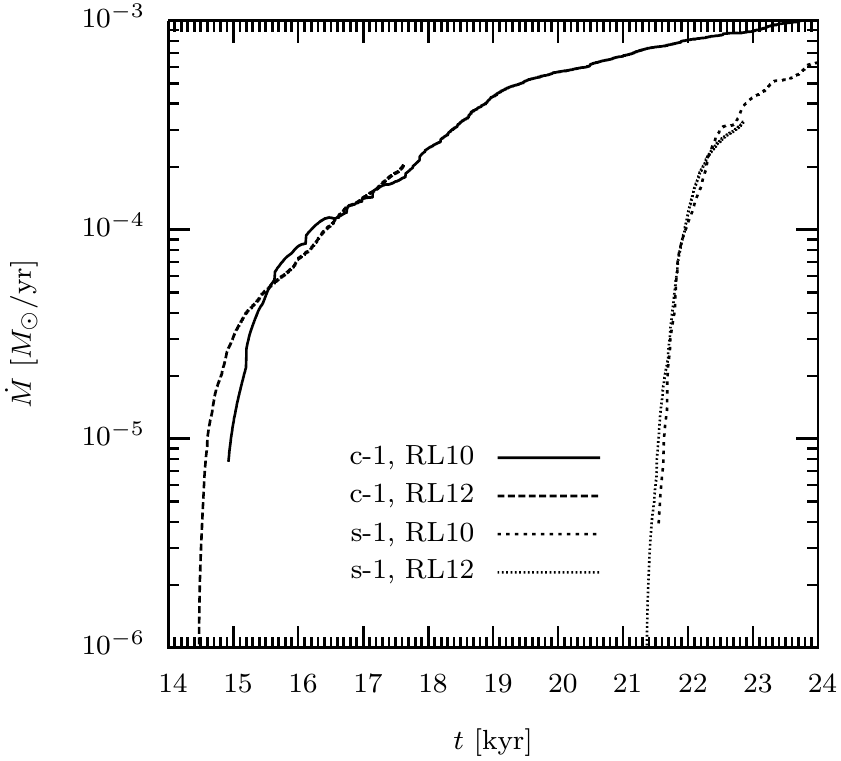}
    \includegraphics[height=8cm]{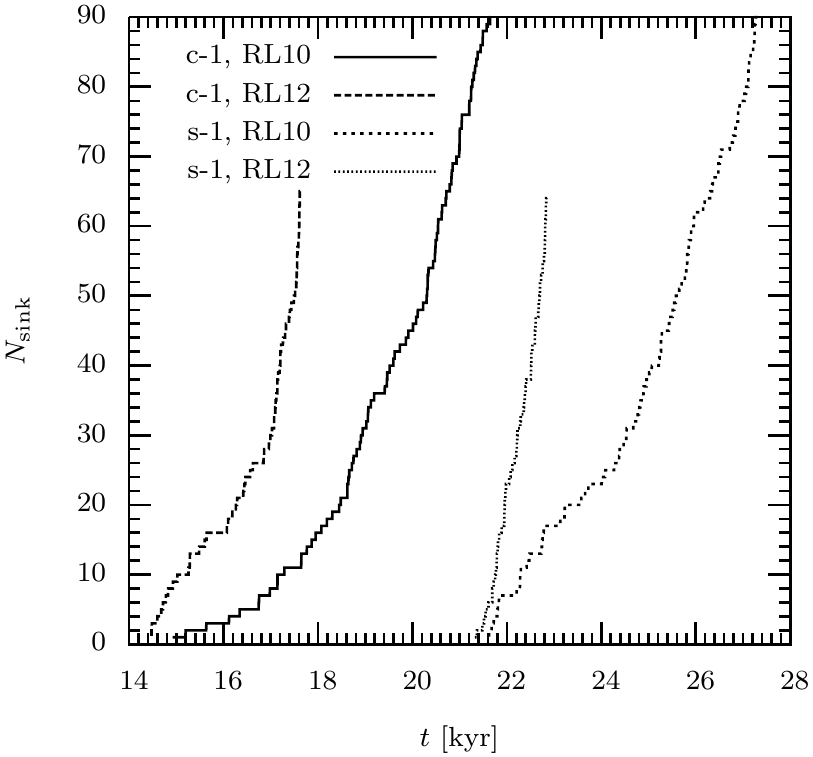}
    \caption{Comparison of the Bonnor-Ebert profiles BE-c-1 and BE-s-1 for resolutions RL10 and RL12. The mass accretion as a function of time (left plot) shows only small differences between the two resolutions. The number of sink particles, however, differs strongly (right plot).}
    \label{fig:BE-resolution-comparison}
  \end{minipage}
\end{figure*}

Due to the flat inner core of this profile, refinement is initiated in a rather large volume of the core, which makes the computational effort for this profile much larger than for the other profiles, and thus the resolution test was done only for a short simulation time. In figure~\ref{fig:BE-resolution-comparison} we compare the total accretion rate, $\dot{M}$, and the number of sink particles $N$ of the Bonnor-Ebert profiles BE-c-1 and BE-s-1 for resolutions RL10 and RL12. The accretion rates are comparable and give roughly the same star formation efficiency with time. However, the number of particles varies significantly with resolution. This is expected, since we use an isothermal equation of state, which does not introduce a physical length scale or density threshold to the problem, i.e., the problem remains scale-free. Changes in the equation of state, in particular if the gas becomes optically thick, will break the scale-free collapse \citep[e.g.,][]{Jappsen05,Krumholz07,Bate09b}.

\subsection{PL20 Profile}
\label{sec:PL20-profile}
\label{sec:scaling-resolution-PL20}
For the concentrated density profile with $\rho\propto r^{-2}$ and the turbulence profile c-1, detailed simulations were run for four different maximum refinement levels: RL07, RL08, RL10, RL12. In all cases only one sink particle was created in the centre of the cloud after a few steps of hydrodynamical evolution.
\begin{figure*}
  \begin{minipage}{180mm}
    \centering
    \includegraphics[height=8cm]{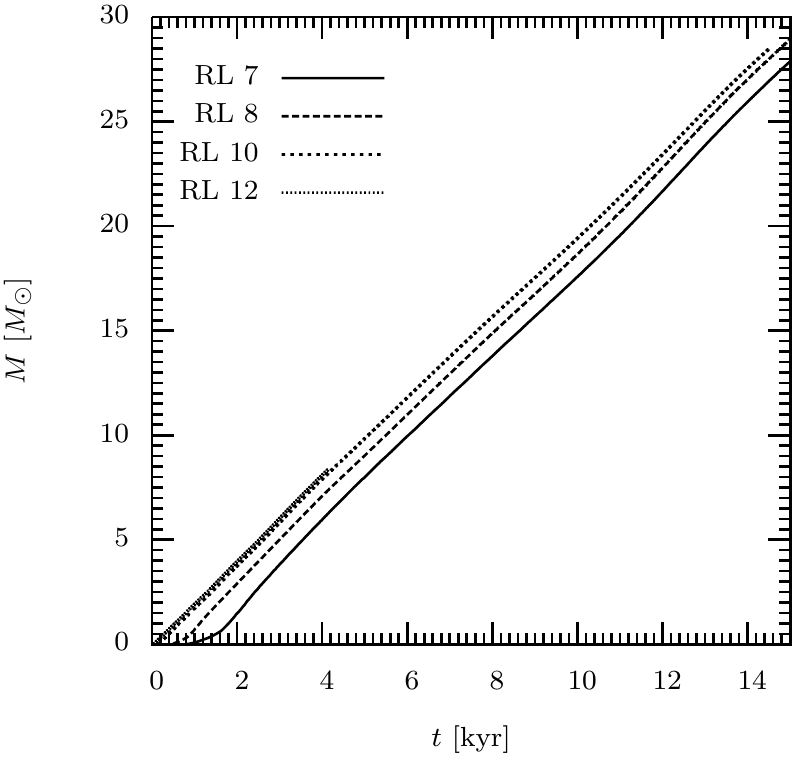}
    \includegraphics[height=8cm]{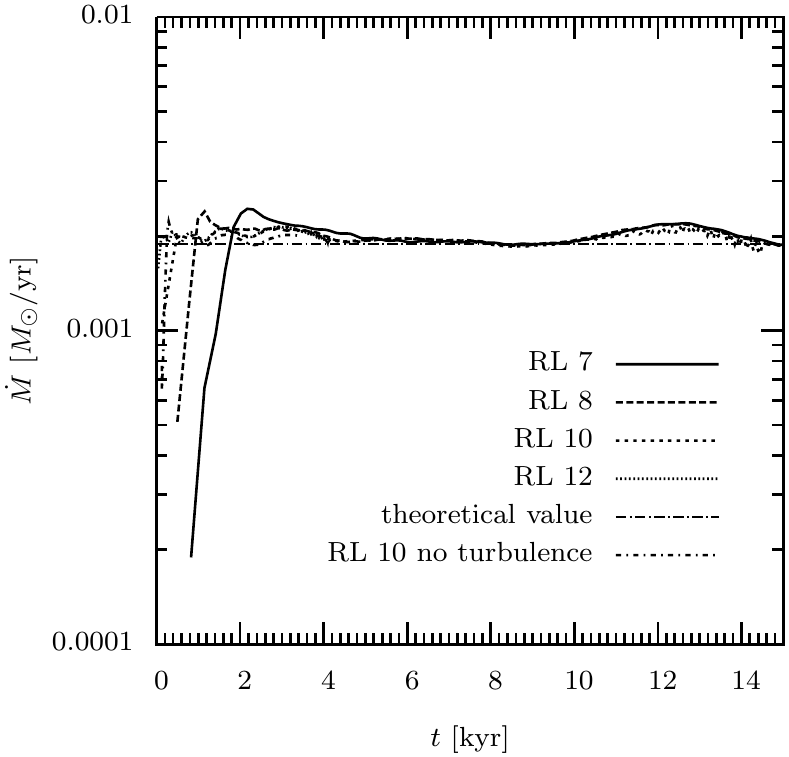}
  \end{minipage}
  \caption{Resolution comparison for the PL20 runs with turbulence field c-1. After an initial evolution time the accretion rates approach the same value for all setups. The differences at the beginning of the simulation are due to different maximum central resolutions. The flattened density function at the centre of the box is much shallower for lower resolutions resulting in larger times for a central collapse.}
  \label{fig:PL20-resolution-comparison}
\end{figure*}

The results for the PL20 runs can be seen in figure~\ref{fig:PL20-resolution-comparison}. The accretion rate onto the protostar $\dot{M}$ does not differ significantly, resulting in the same slope of the mass $M$ as a function of time. The different evolution of the accretion rate at the very beginning of the simulation is due to the different geometrical setup conditions (see sec.~\ref{sec:PLs}). The larger size of the smallest cell for lower refinement levels results in a much coarser density distribution in the centre of the cloud and needs more evolution time in order to develop a sink particle with constant accretion rate. The theoretical value for the accretion rate fits the simulated values very well (see sec.~\ref{sec:PL20-self-similarity}). The comparison with a simulation without turbulent velocities only shows minor differences.

\section{Tidal Forces}
\label{sec:tidal_forces}
The tidal acceleration in a spherically symmetric setup at distance $r$ from the centre with an enclosed mass $M$ is given by
\begin{equation}
  a_\text{tidal}(r) = G \frac{M}{(r\pm\Delta r)^2},
\end{equation}
where $G$ is the gravitational constant and $\Delta r \ll r$. The enclosed mass can then considered to be constant within the variation $\Delta r$. Given a density profile of the form $\rho(r)\propto r^{-p}$ yields a mass function $M(r)\propto r^{3-p}$, and the tidal acceleration scales as
\begin{equation}
  a_\text{tidal}(r) \propto r^{1-p}.
\end{equation}
The derivative with respect to $r$,
\begin{equation}
  \frac{\partial a_\text{tidal}}{\partial r}(r) \propto (1-p)\,r^{-p},
\end{equation}
changes sign at $p=1$. For $p<1$, $a_\text{tidal}$ increases with radius ($\partial a_\text{tidal}/\partial r > 0$) and therefore compresses material at radius $r$. For $p>1$, $\partial a_\text{tidal}/\partial r < 0$ and shears condensations apart.

\end{appendix}

\end{document}